\documentclass[prd,
floatfix,
superscriptaddress,
nofootinbib,
amsmath,amssymb,
twocolumn,
longbibliography
]{revtex4-2}
\usepackage{slashed}
\usepackage{booktabs}
\usepackage{graphicx}
\usepackage[newitem,newenum]{paralist}
\usepackage{threeparttable}
\usepackage[utf8]{inputenc}
\usepackage[%
separate-uncertainty=true,%
print-zero-exponent=false,% whether to put n x 10^0 instead of n
print-unity-mantissa=false% whether to put 1 x 10^n instead of 10^n
]{siunitx}
\usepackage[dvipsnames]{xcolor}
\usepackage{bbm}
\usepackage[colorlinks,allcolors=blue,linktocpage]{hyperref}
\usepackage{ragged2e} % Provides \justifying
\usepackage{braket}
\usepackage[caption=false]{subfig}
\usepackage[margin=false,inline=true,index=true,final]{fixme}
\usepackage{graphicx}
\fxsetup{inlineface=\footnotesize}
\fxusetheme{colorsig}
\FXRegisterAuthor{oleg}{anoleg}{O.R.}
\FXRegisterAuthor{mads}{anmads}{M.H}
\FXRegisterAuthor{edis}{anedis}{E.T.}
\FXRegisterAuthor{rikke}{anrikke}{R. S. K.}
\usepackage{mathrsfs}
\newcommand{\delphi}{\text{\tiny\scshape delphi}}
\newcommand{\mdv}{m_{\text{\scshape dv}}}
\newcommand{\rdv}{r_{\text{\scshape dv}}}
\newcommand{\Ut}[1]{\Theta_\tau^{#1}}
\newcommand{\Br}{\mathop{\mathrm{Br}}}
\newcommand{\parfrac}[2]{\left(\frac{#1}{#2}\right)}
\newcommand{\vvbar}[1]{\overset{{\scriptscriptstyle(-)}}{\nu}_{#1}}

\usepackage[capitalise]{cleveref}

\usepackage{tikz}
\usetikzlibrary{shapes,arrows,positioning,automata,backgrounds,calc,er,patterns}
\usepackage[compat=1.1.0]{tikz-feynman}

\begin{document}
\title{Constraining Heavy Neutral Leptons Coupled to the Tau-Neutrino Flavor\newline%
at the Large Hadron Collider}
\author{Edis Devin Tireli}
\email{edis.tireli@sund.ku.dk}
\affiliation{Niels Bohr Institute, University of Copenhagen, Blegdamsvej 17, DK-2100 Copenhagen, Denmark}
\affiliation{Department of Neuroscience, University of Copenhagen, Blegdamsvej 3B, DK-2200 Copenhagen, Denmark}

\author{Rikke Stougaard Klausen}
\affiliation{Niels Bohr Institute, University of Copenhagen, Blegdamsvej 17, DK-2100 Copenhagen, Denmark}

\author{Oleg Ruchayskiy}
\affiliation{Niels Bohr Institute, University of Copenhagen, Blegdamsvej 17, DK-2100 Copenhagen, Denmark}
\begin{abstract}
Displaced vertex (DV) signatures at colliders offer a powerful probe of new long-lived particles beyond the Standard Model. Among the best-motivated candidates are \emph{heavy neutral leptons} (HNLs) -- heavier counterparts of Standard Model neutrinos -- which can account for the origin of neutrino masses and potentially produce di-leptonic DV signatures.

In this study, we demonstrate how existing DV searches at the LHC can be extended to probe HNLs that couple predominantly to the tau-neutrino flavor. While current search strategies rely on identifying a prompt lepton alongside a displaced vertex, we show that analyzing events \emph{without} a prompt lepton enables sensitivity to the process $pp \to W \to \tau N$, where the tau decays hadronically and the HNL subsequently decays to a lepton pair and a neutrino.

We perform detailed Monte Carlo simulations of this process with HNLs decaying to $\mu^+\mu^-$ or $e^+e^-$ final states, apply ATLAS-inspired selection criteria, and optimize signal sensitivity. In particular, we demonstrate that appropriate cuts in the plane of di-lepton invariant mass  and DV radial position  significantly enhance signal visibility.
We propose several such optimized strategies and show that even with Run~2 data ($\SI{139}{\per\femto\barn}$), existing bounds can be improved by more than an order of magnitude. Future high-luminosity runs may strengthen sensitivity by up to three orders of magnitude compared to current limits.
\end{abstract}
\maketitle
\section{Introduction}
Right-handed (sterile) neutrinos have long been proposed as an explanation for neutrino oscillations.
Over times it has been understood, that the same particles can lead to generation of baryon asymmetry the Universe \cite{Davidson:2008bu,Pilaftsis:2009pk,Klaric:2021cpi}, provide a  dark matter candidate, see e.g.,  \cite{Boyarsky:2018tvu} for review,
or account for all of these phenomena \cite{Asaka:2005an,Asaka:2005pn,Boyarsky:2009ix}.
While neutrino oscillations do not constrain the mass of right-handed neutrino, the requirement of successful baryogenesis limits the mass from below at GeV level~\cite[see e.g.,][]{Klaric:2021cpi,Bondarenko:2021cpc}
This opens an exciting possibility that particles, responsible for the major beyond-the-Standard-Model (BSM) phenomena, can be searched for in laboratories.

The search for these particle is challenging, as right-handed neutrinos do not participate in the gauge interactions of the Standard Model (SM).
They couple to the left-handed lepton doublets through Yukawa interactions with the Higgs field.
Their neutral nature also allows them to acquire a Majorana mass independently of the SM Higgs mechanism.
When transitioning from the flavor basis to the mass basis, Yukawa mixing gives rise to both light and heavy mass states.
The active neutrinos acquire small masses, consistent with observations, while the heavier states inherit weak-like interactions, albeit with a highly suppressed strength.
This suppression is characterized by the matrix of \emph{mixing angles}, $\Theta_{\alpha I}$, proportional to Yukawas $Y_{\alpha I}$ and inversely proportional to masses $M_I$.\footnote{Here and below, $\alpha$ denotes the lepton flavor index, while $I=1,2,\dots$ labels right-handed (sterile) states.}
The heavy mass states are known as \emph{heavy neutral leptons} (HNLs).
Due to their suppressed interactions, HNLs couple only feebly to ordinary matter, making them difficult to detect in experiments.

To account for neutrino oscillations, at least two HNLs are required. Their masses must be nearly degenerate, and their mixing angles related by a phase, $\Theta_{\alpha 1} \simeq \pm i \Theta_{\alpha 2}$, in order to keep the light neutrino masses small and suppress large radiative corrections~\cite{Pilaftsis:1991ug, Shaposhnikov:2006nn, Kersten:2007vk, Tastet:2019nqj, Fernandez-Martinez:2022gsu}. Under these conditions, current neutrino oscillation data constrains the ratio of the mixing angles, restricting the viable regions of HNL parameter space; see~\cite{Tastet:2021vwp, Bondarenko:2021cpc} for reviews. Phenomenologically, this setup is equivalent to a single HNL with a common mass $m_N$ and total mixing angle $\Theta^2 = \sum_{\alpha, I}|\Theta_{\alpha I}|^2$.

\begin{figure}[!b]
    \centering
    \begin{tikzpicture}
    \begin{feynman}
    \vertex (a);
    \vertex [left=of a](i1) {\(W^{+}\)};
    \vertex [below right=of a](m1);
    \vertex [above right=of a](f1) {\(\tau^{+}\)};
    \vertex [below right=of a] (b);
    \vertex [below right=of b] (f2) {\(\overset{\textbf{\fontsize{3pt}{3pt}\selectfont(---)}}{\nu}_{\tau}\)};
    \vertex [right=of b] (m2);
    \vertex [right=of b] (c);
    \vertex [above right=of c] (f3) {\(\ell^{-}\)};
    \vertex [below right=of c] (f4) {\(\ell^{+}\)};
    \diagram* {
    (i1) -- [boson] (a),
    (a) -- [anti fermion] (f1),
    (a) -- [scalar, edge label'=\(N\)] (m1),
    (b) -- [fermion] (f2),
    (b) -- [boson, edge label=\(Z^{*}\)] (m2),
    (c) -- [fermion] (f3),
    (c) -- [anti fermion] (f4),

    };
    \end{feynman}
\end{tikzpicture}
\caption{\justifying\textbf{Main process.} Feynman diagram~\protect\eqref{1} showing the production of the HNL via the mixing with the $\tau$-flavor and its subsequent decay mediated by the neutral current interaction and governed by the same mixing angle.}
    \label{fig:feynman}
\end{figure}
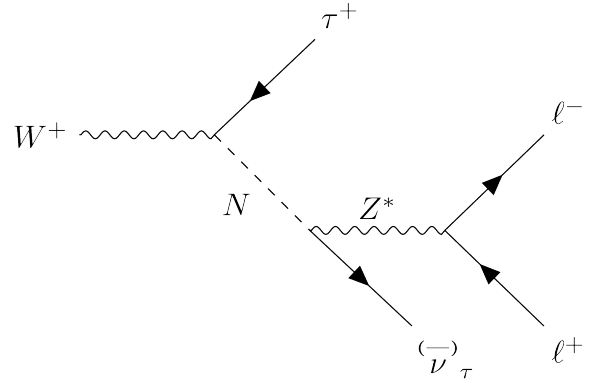

Feeble strength of interaction may result in HNLs being \emph{long-lived particles}, which significantly impacts their search strategies. Depending on their mass and mixing angles, HNLs can travel macroscopic distances before decaying, making them prime candidates for fixed target experiments, such as e.g., SHiP \cite{Alekhin:2015byh, SHiP:2018xqw} or displaced vertex searches at high-energy colliders~\cite{CMS:2018iaf,ATLAS:2019kpx,LHCb:2020wxx,ATLAS:2022atq,CMS:2022fut,CMS:2024xdq,CMS:2024ake,ATLAS:2025uah}. In addition to collider experiments, HNLs are also actively searched in wide variety of particle physics experiments, and even through astrophysical observations all of which probe different regions of the HNL parameter space \cite{Abdullahi_2023}.

Unlike other leptons, HNL do not exhibit flavour universality and therefore each $\Theta_\alpha$ should be measured independently, either through direct experimental searches (see below) or indirectly \cite{Urquia-Calderon:2022ufc,Blennow:2023mqx}.

Strategy and sensitivity of the measurements depends on the flavor.
Collider searches have been least sensitive to the mixing with the $\tau$-flavour owing to the difficulty of reconstructing $\tau$-leptons in the final state.
As a result, the coupling with the $\tau$-flavour has never been directly explored by the LHC collaborations; only its combinations with other flavours have been studied \cite{Arnau_MSc, CMS:2024xdq} or constraint indirectly, assuming a particular ratio between flavours, c.f., \cite{ATLAS:2022atq}, motivated by neutrino oscillations.
Other recent searches include \cite{BaBar:2022cqj,Belle:2024wyk} and  \cite{IceCube:2025kve}, see also \cite{Boiarska:2021yho} for reinterpretation of the previous experimental results.

\bigskip

\emph{In this work we demonstrate that existing data and analysis pipelines actually allow to probe the mixing of HNLs with the 3rd generation at the LHC.}
Namely, we consider the process
\begin{equation}
  \label{1}
  p \, p \to \tau^\pm N, \quad N \to \ell^+ \ell^- \vvbar{}
\end{equation}
where $N$ is HNL, leptons $\ell\in\{e, \mu\}$ and $\vvbar\tau$ is neutrino or anti-neutrino of $\tau$-flavor.
The Feynman diagram for the process is shown in Figure~\ref{fig:feynman}.
The probability of this process is proportional to $\Theta_\tau^4$ and therefore the study of this signal gives a direct access to the coupling of the 3rd generation neutrinos with HNL.

\begin{figure}[!t]
    \centering
    \includegraphics[width=\linewidth]{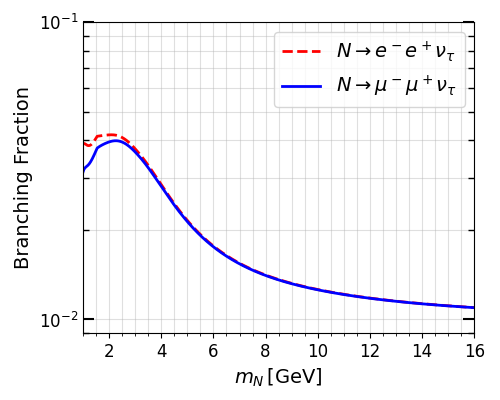}
    \caption{\justifying%
    The branching ratios of an HNL decaying into a lepton pair and $\nu_\tau$ for the flavor mixing pattern $(\Theta_e^2, \Theta_\mu^2, \Ut2) = (0,0,1)$. The branching ratios are independent of the specific value of $\Ut2$.}
    \label{fig:branching_ratio}
\end{figure}

A simple estimate demonstrates the viability of these searches.
The HNL production cross-section at $\sqrt s = \SI{13}{TeV}$ is obtained by rescaling by $\Ut2$ the measured cross-section $\sigma(pp \to W)\Br(W\to \ell\nu) = \SI{20.6}{nb}$~\cite{ATLAS:2016fij}.
The branching ratio, $\Br(N \to \mu^+ \mu^- \nu_\tau/\bar{\nu}_\tau)$~\cite{Bondarenko:2018ptm} is in the range of $1-2\%$ for the masses of interest (see Figure~\ref{fig:branching_ratio}).
The number of lepton pairs, produced via HNL decay is given by
\begin{equation}
  \label{eq:estimate_analytic}
  N_{\ell\ell} = \mathcal{L} \cdot \Theta_{\tau}^2 \cdot \sigma(pp \to \tau_{\text{h}} N)\cdot \text{Br}(N\rightarrow \ell \ell \nu_\tau) \cdot  \epsilon_{\text{acc}}
\end{equation}
where $\epsilon_{\text{acc}}$ is the signal acceptance,  $\tau_{\text{h}}$ refers to $\tau$-leptons decaying either hadronically and $\mathcal L$ is the integrated luminosity.
Pluging in the number this gives
\begin{widetext}
\begin{equation}
      \label{eq:estimate}
  N_{\ell\ell} \simeq 10\parfrac{\mathcal L}{\SI{300}{\femto\barn^{-1}}}\parfrac{\Br}{\num{1.2e-2}}\parfrac{\Ut2}{{\Theta_{\tau}^2}_{\delphi}}\parfrac{\strut\epsilon_{\text{acc}}}{0.01}
\end{equation}
\end{widetext}
where we normalized the result to the DELPHI limit \cite{DELPHI:1996qcc} ${\Theta_{\tau}^2}_{\delphi}\simeq \num{1.3e-5}$.
For HNL traveling macroscopic distances from the interaction point, the SM background is heavily suppressed and therefore the searches may be viewed as background free \cite{Bondarenko:2019tss,Drewes_2020}.
The estimate~(\ref{eq:estimate}) indicates that with an acceptance rate of around $\epsilon_{\text{acc}} \gtrsim 1\%$, we can probe a previously unexplored region of the HNL parameter space, motivating the current analysis.

The paper is organized as follows.
We begin with presenting our results -- potential exclusion regions for searches in $\mu\mu$, $ee$ or combined channels and for different luminosities $\mathcal{L} = 139,\, 300,\, 3000\si{\per\femto\barn}$ (Section~\ref{sec:result}).
We then move to the description of our method and the assumptions underlying the displaced vertex selection criteria (Section~\ref{sec:analysis}).
We begin by outlining the general event selection (Section~\ref{subsec:selection}), including the treatment of the decay volume (Section~\ref{sec:decay_volume}) and the invariant mass selection (Section~\ref{subsubsec:invmass}).
Special care is taken to address the role of prompt $\tau$-leptons in the signal topology (Section~\ref{sec:tau_analysis}).
Background considerations and the strategies employed to suppress them are discussed in Section~\ref{sec:background}.
Finally, we summarize our findings and their implications in Section~\ref{sec:conclusion}.

\section{Results}
\label{sec:result}

We begin by presenting our main results---\emph{signal sensitivity}.
Specifically, we identify regions in the parameter space where we expect $N \geq 3$ signal events. This criterion is motivated by the fact that the searches are predominantly background-free (as we will argue below), making the $N = 3$ boundaries equivalent to $95\%$ confidence level bounds.

The analysis is performed at generator level---we generate the Monte Carlo signal sample and apply the selection criteria listed in Table~\ref{tab:cut_table}. While our kinematic cuts are similar to the signal selection criteria used in the ATLAS analysis~\cite{ATLAS:2022atq}, we extend the exploration of cuts in the $\mdv$ and $\rdv$ plane beyond those considered in the ATLAS study. Specifically, we investigate two approaches (see Section~\ref{subsubsec:invmass} below):
\begin{compactitem}[--]
    \item \textbf{Flat selection:} The invariant mass criterion is flat, with $m_N \geq \SI{5}{GeV}$ for both muon pairs and electron-positron pairs, across the whole fiducial volume.
    \item \textbf{Piecewise selection:} The invariant mass criterion depends on the position inside the fiducial volume. This approach is motivated by the structure of the background observed in~\cite{PhysRevLett.131.061803}; see also~\cite{Appelt:2024esk}, where a similar background veto was applied.

\end{compactitem}

The analysis is performed for two channels: $e^+e^-$ and $\mu^+\mu^-$. Since an HNL with predominant $\tau$-coupling decays into both channels with approximately equal probability, we also consider a \textit{combined} ($e^+e^- + \mu^+\mu^-$) channel.

\begin{figure*}[p]
    \centering
    \subfloat[]{\includegraphics[width=0.45\textwidth]{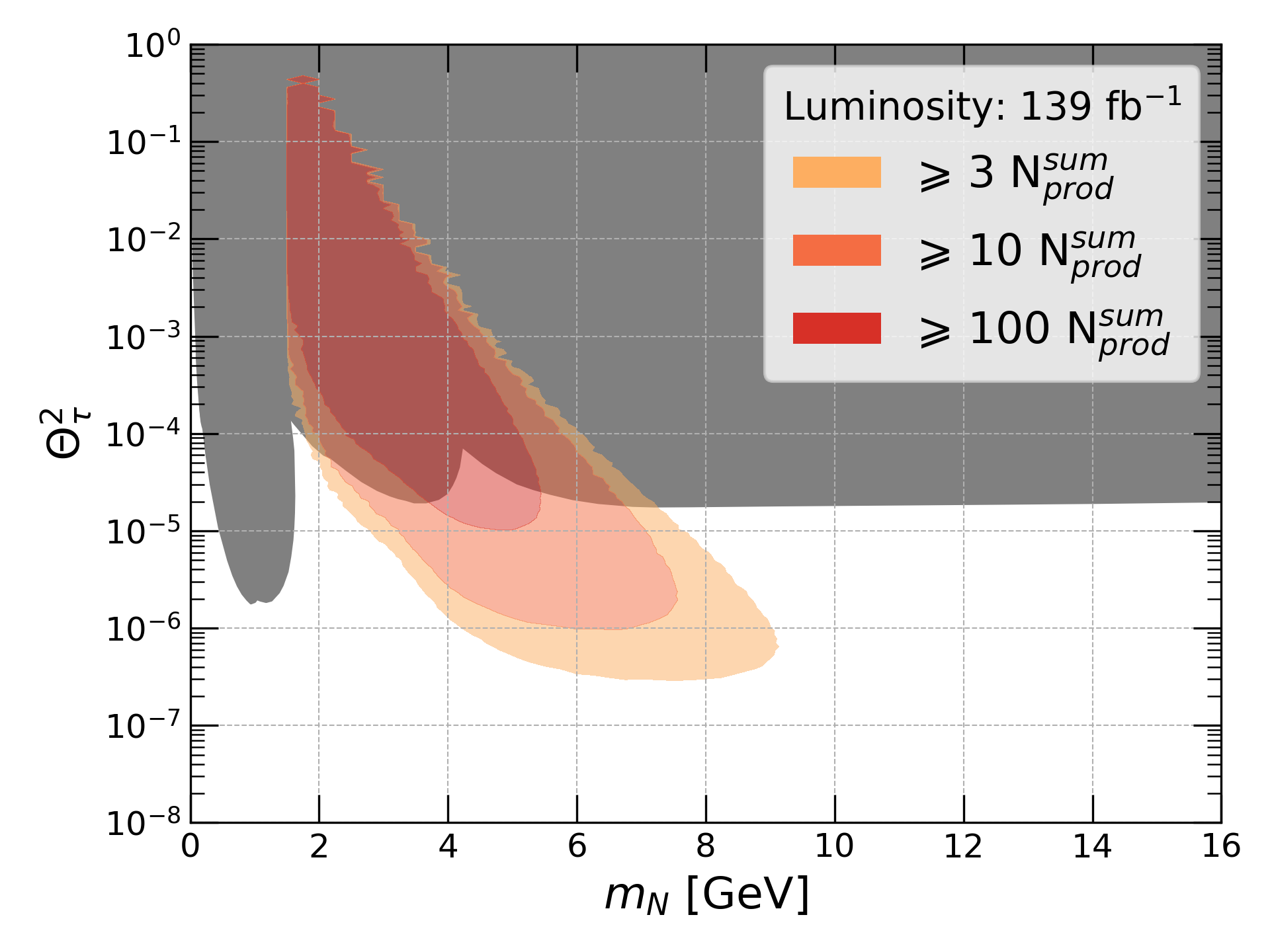}}~
    \subfloat[]{\includegraphics[width=0.45\textwidth]{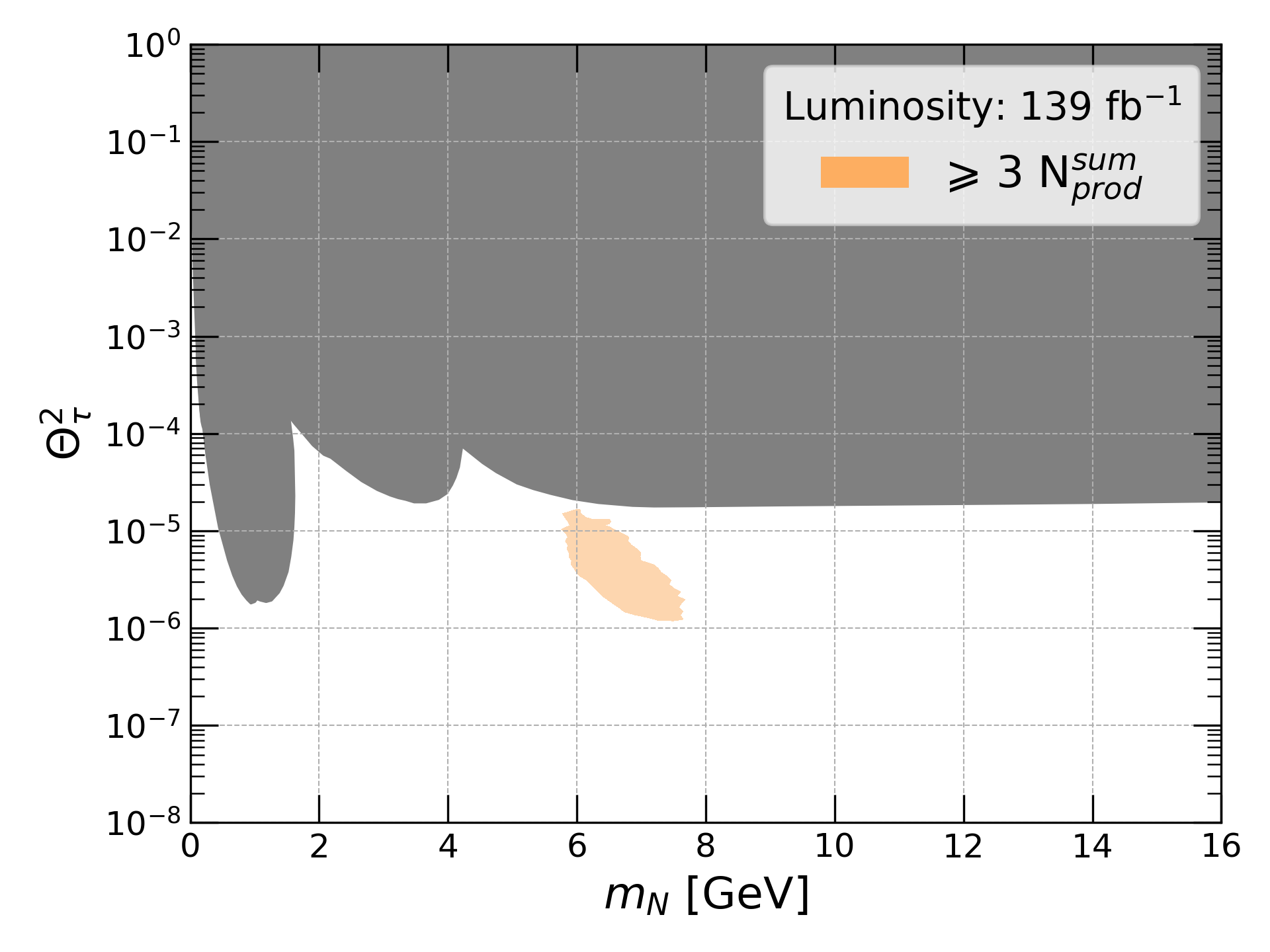}}\\
    \subfloat[]{\includegraphics[width=0.45\textwidth]{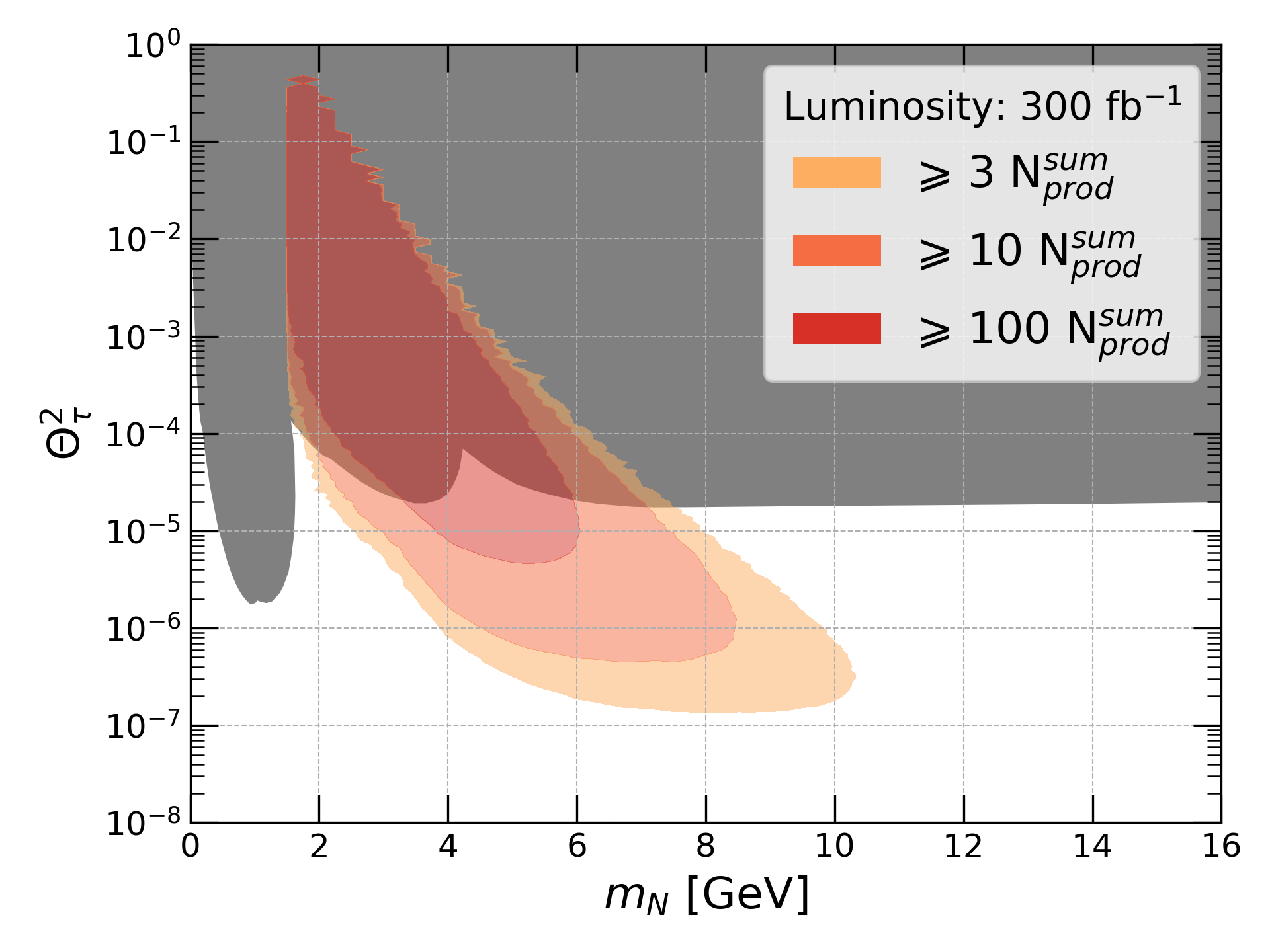}}~
    \subfloat[]{\includegraphics[width=0.45\textwidth]{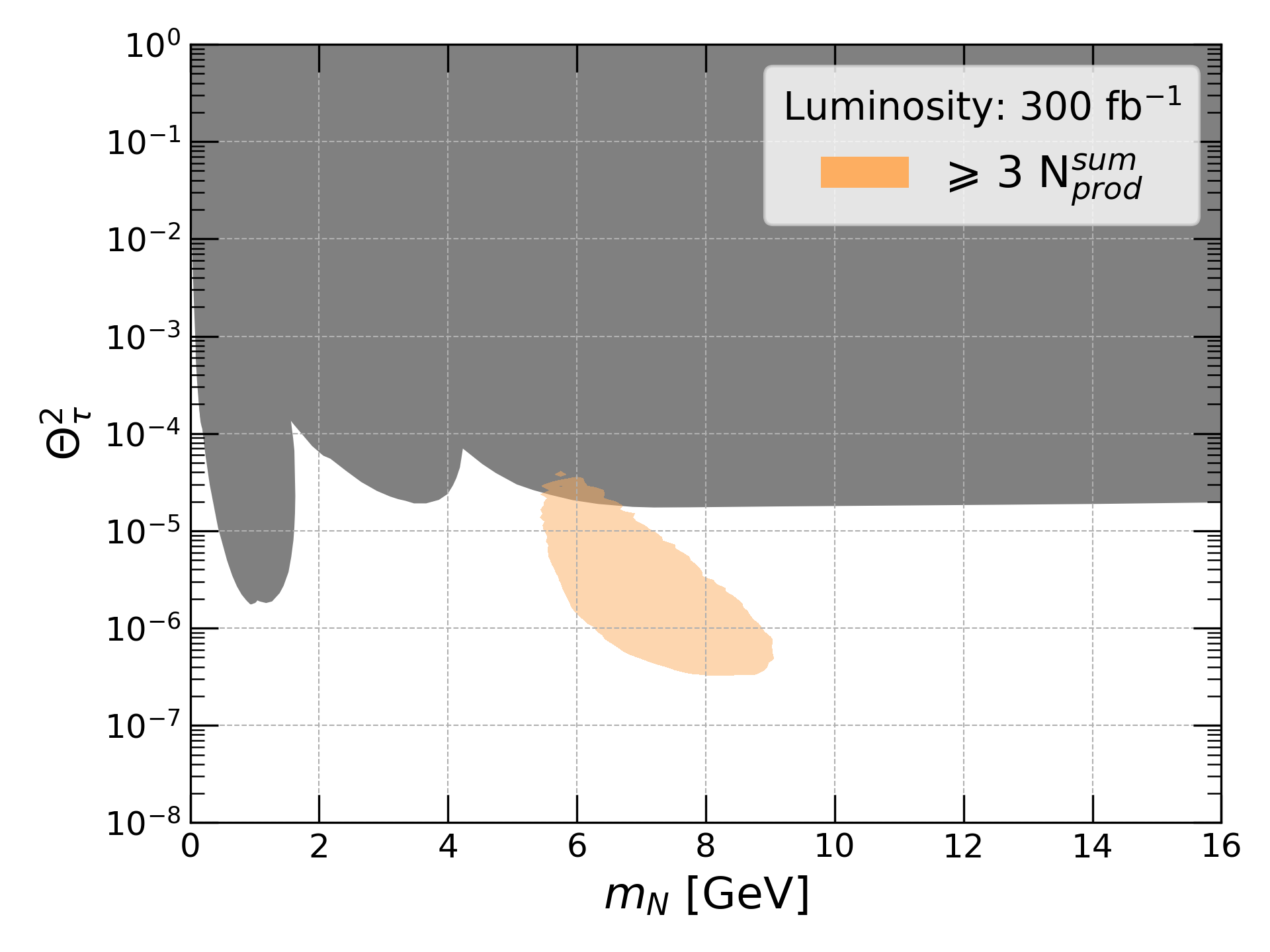}}\\
    \subfloat[]{\includegraphics[width=0.45\textwidth]{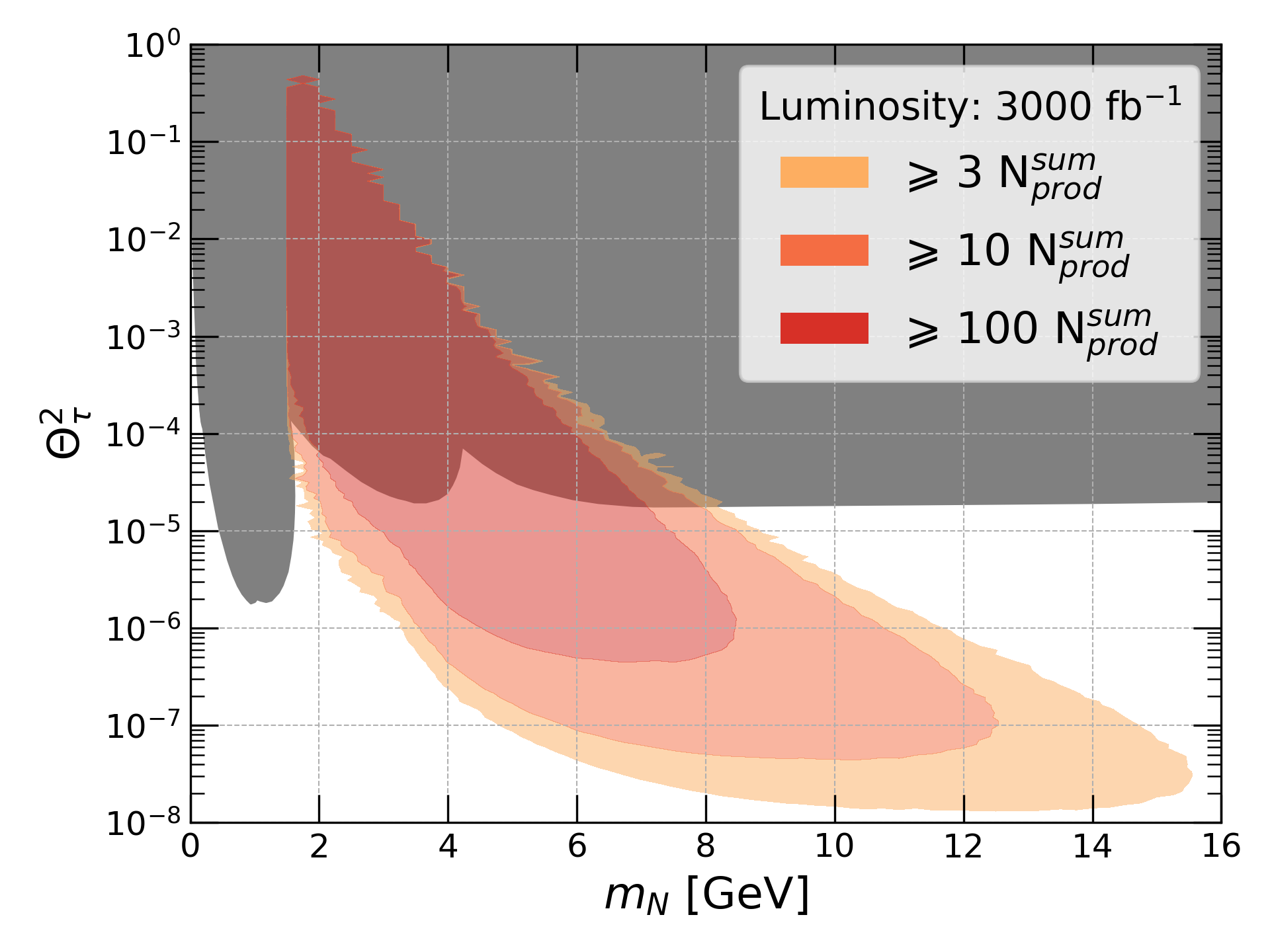}}~
    \subfloat[]{\includegraphics[width=0.45\textwidth]{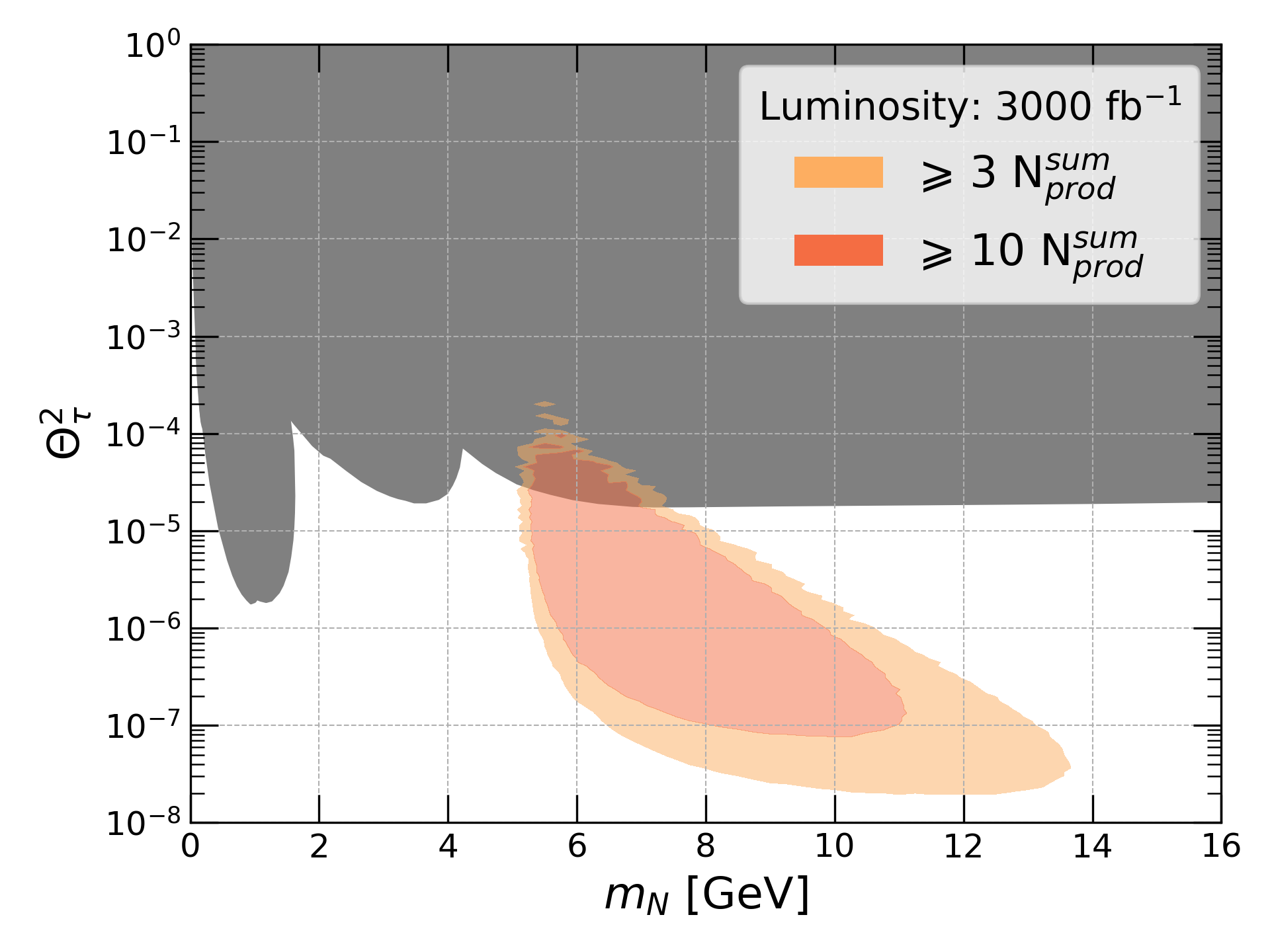}}
    \caption{\justifying{}\textbf{Sensitivity of the combined channel as a function of the number of observed events.} The signal sensitivity (for 3, 10, or 100 expected events) in the combined $e^{+}e^{-} + \mu^{+}\mu^{-}$ channel at integrated luminosities of $\SI{139}{\per\femto\barn}$ (top), $\SI{300}{\per\femto\barn}$ (middle), and $\SI{3000}{\per\femto\barn}$ (bottom). The left column (panels a, c, and e) corresponds to the \textbf{piecewise} invariant mass selection, while the right column (b, d, and f) corresponds to the \textbf{flat} invariant mass selection. Notably, in the flat selection case, no regions with $100$ events are observed (panel f), and even the $10$-event regions disappear in panels b and d, highlighting the critical role of the DV cut choice.}
    \label{fig:main_result_Nevents}
\end{figure*}

Figure~\ref{fig:main_result_Nevents} illustrates the sensitivity for the combined channel, assuming \( 3 \), \( 10 \), or \( 100 \) signal events at various luminosities. %The \( N \geq 3 \) events correspond to the \( 95\% \) confidence level in the absence of a significant Standard Model (SM) background.
The cases $N = 10$ and $100$ demonstrate how the sensitivity degrades should background contributions be present.
One can see that a significant fraction of previously unexplored parameter space can be probed already with data from Run~2 luminosity ($\mathcal L = \SI{139}{fb^{-1}}$).
The high-luminosity run may allow to probe up to 3 orders of magnitude deep for masses around $\SIrange{10}{15}{GeV}$, as compared to the current state-of-the-art.

\begin{figure*}[p]
    \centering
    \subfloat[]{\includegraphics[width=0.45\textwidth]{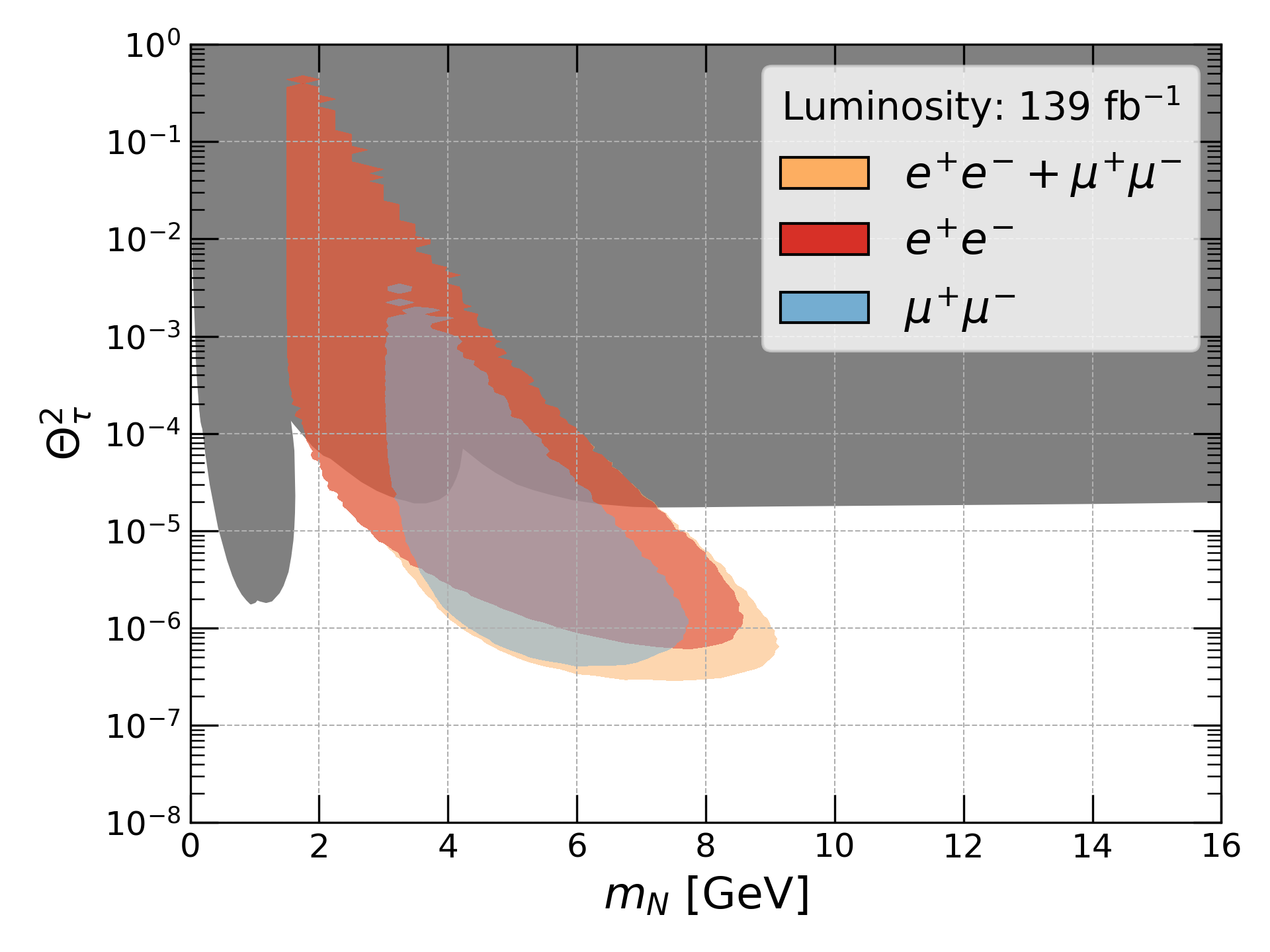}}~
    \subfloat[]{\includegraphics[width=0.45\textwidth]{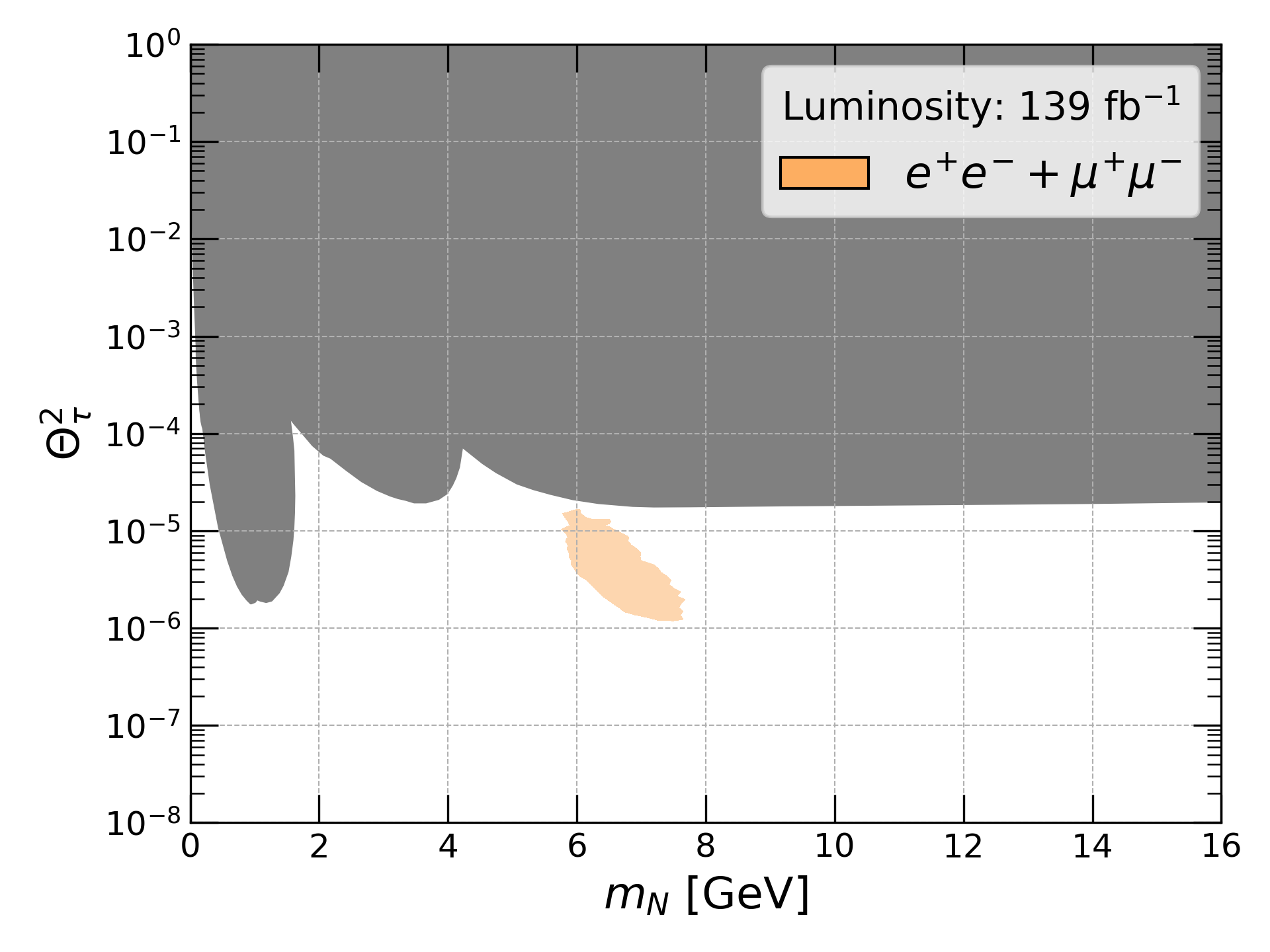}}\\
    \subfloat[]{\includegraphics[width=0.45\textwidth]{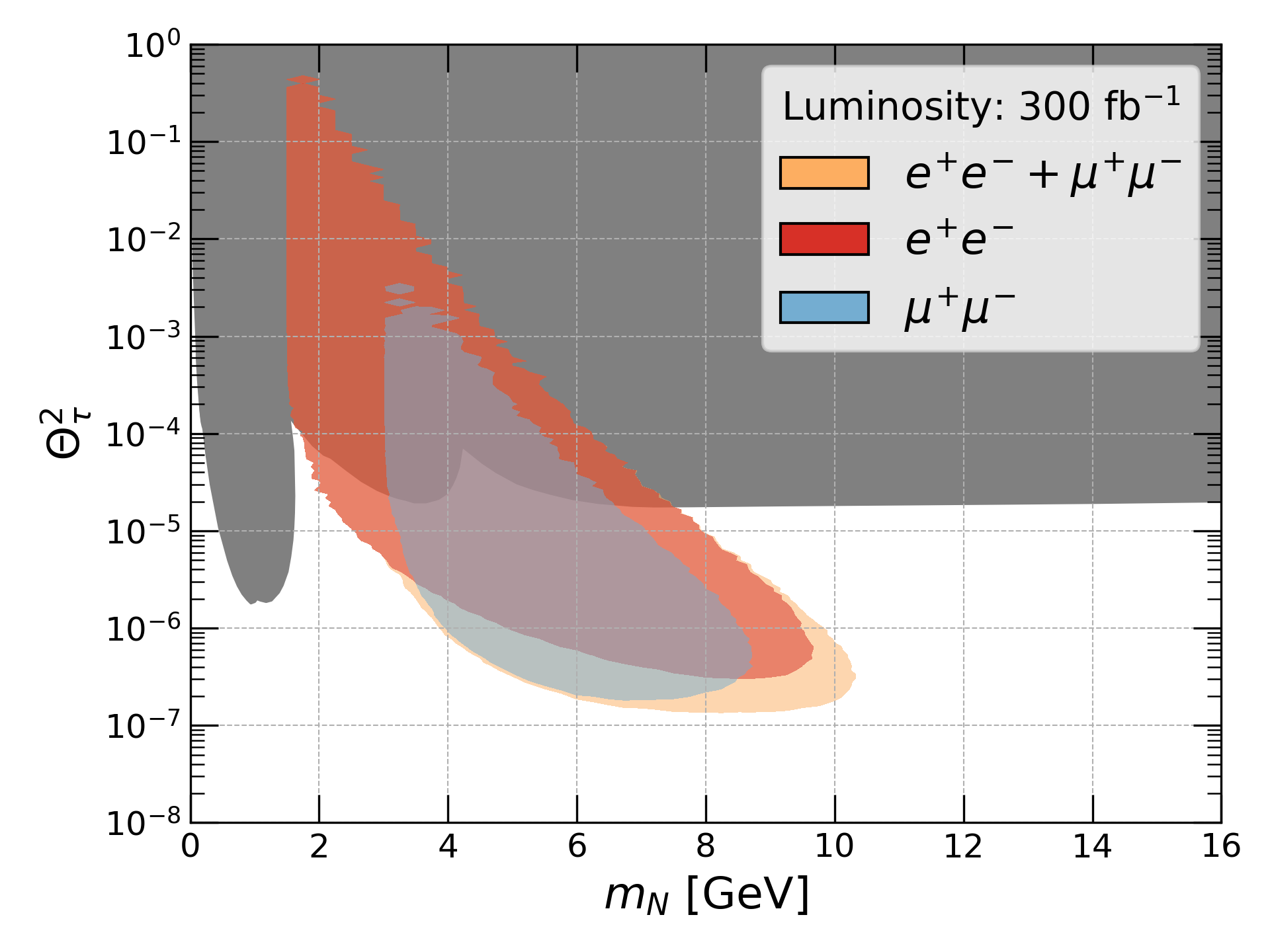}}~
    \subfloat[]{\includegraphics[width=0.45\textwidth]{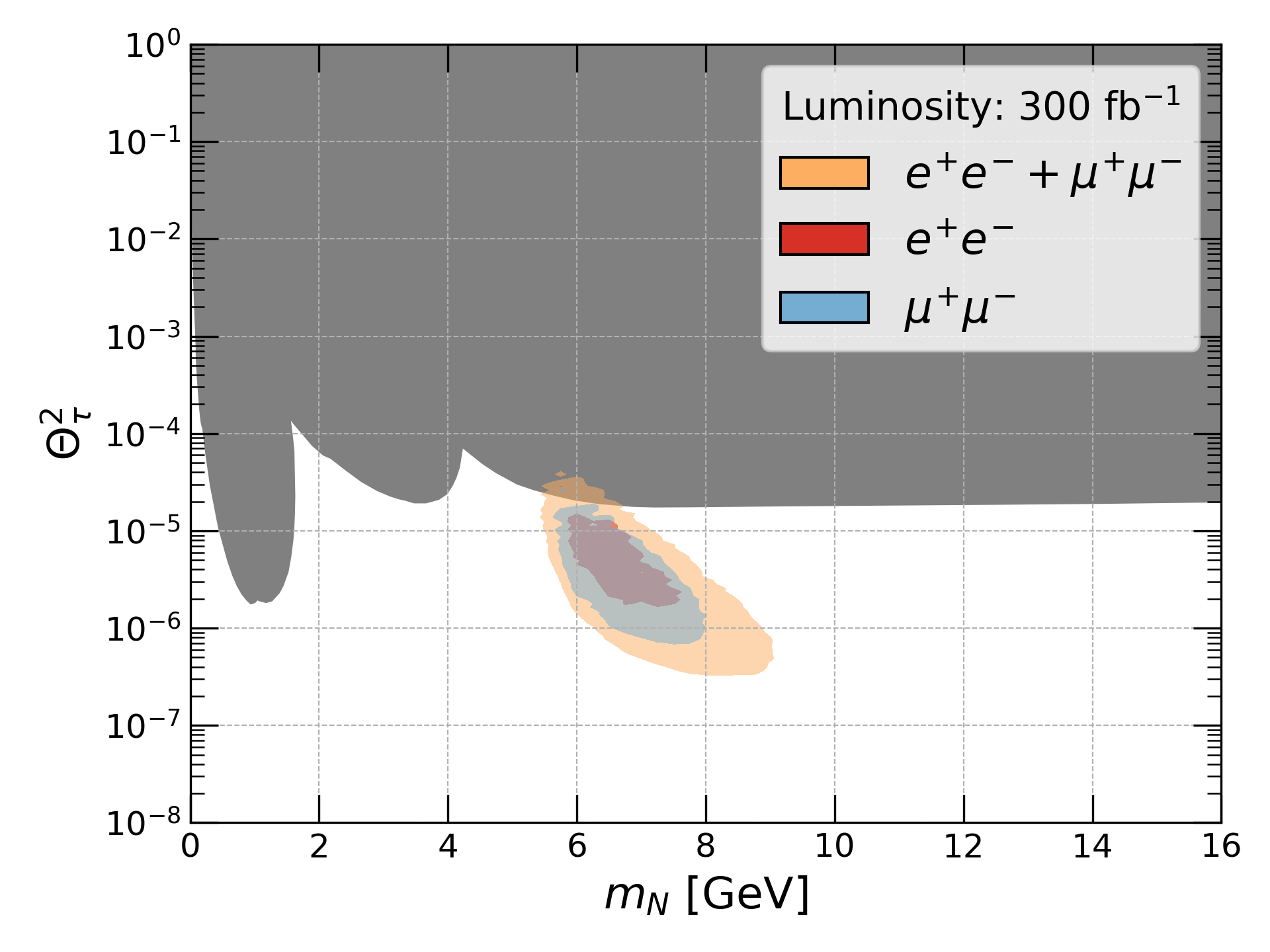}}\\
    \subfloat[]{\includegraphics[width=0.45\textwidth]{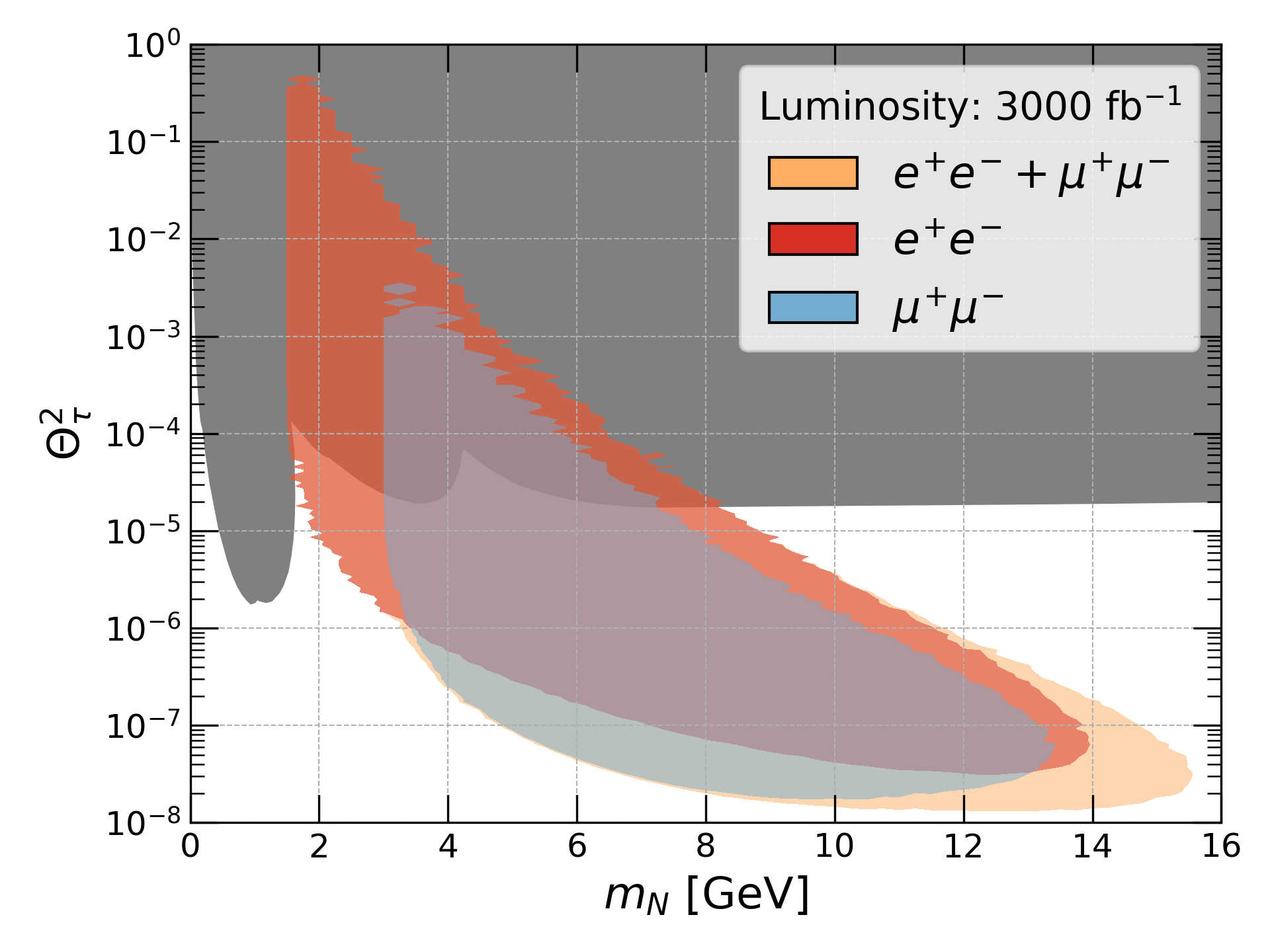}}~
    \subfloat[]{\includegraphics[width=0.45\textwidth]{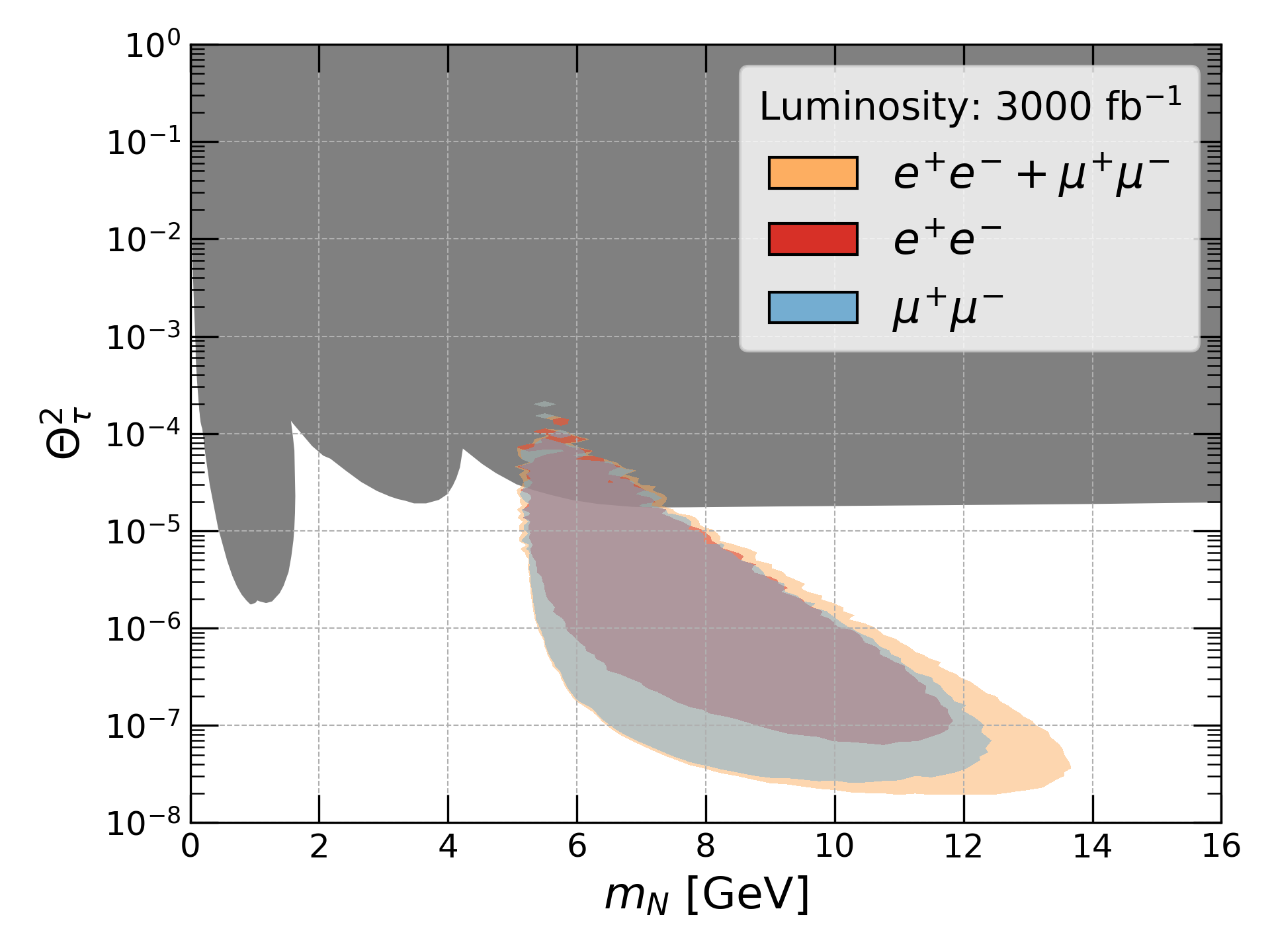}}
    \caption{\justifying{}The signal sensitivity for $N \geq 3$ events of the $e^{+}e^{-}$ (red), $\mu^{+}\mu^{-}$ (blue), and the combined $e^{+}e^{-} + \mu^{+}\mu^{-}$ (yellow) channel at 139 fb$^{-1}$ (top), 300 fb$^{-1}$ (middle) and 3000 fb$^{-1}$ (bottom). The left column (a,c, and e) is the \textbf{piecewise} invariant mass cut, and the right column (b, d, and f) is the \textbf{flat} invariant mass cut.}
    \label{fig:main_result_channels}
\end{figure*}

Figure~\ref{fig:main_result_channels} further details how electron and muon channels contribute to the resulting sensitivity at various luminosities for two different choices of $\mdv,\rdv$ cuts.
One can see, in particular, that for $\mathcal L = \SI{139}{fb^{-1}}$ and flat cut (Figure~\ref{fig:main_result_channels}.b) none of the individual channels contribute 3 signal events, while their combination does.
For all other luminosities/cuts combinations both channels contribute individually and largely overlap, making this a powerful way of rejecting combinatorial backgrounds.

The grey area on the plot represents regions previously explored by experiments with the lowest sensitivity such as DELPHI~\cite{DELPHI:1996qcc} and BEBC~\cite{Barouki:2022bkt}.


%\FloatBarrier

\begin{table*}[t]
\centering
\caption{Selection Criteria for Muon and Electron Channels}
\label{tab:selection_criteria}
\begin{threeparttable}
\begin{tabular}{@{}>{\raggedright}p{7cm}rr@{}}
\toprule
\textbf{Criteria} & \textbf{Muon Channel} & \textbf{Electron Channel} \\
  \midrule
  No hard prompt leptons & $p_T^{\mathrm{prompt}} < \SI{3}{GeV}$ & $p_T^{\mathrm{prompt}} < \SI{4.5}{GeV}$ \\
Minimum transverse momentum $p_T^{\mathrm{min}}$ & 3 GeV & 4.5 GeV \\
Minimum invariant mass $M_{\mathrm{inv}}^{\mathrm{min}}$ & Piecewise function\tnote{\dag} or 5 GeV & Piecewise function\tnote{\dag} or 5 GeV \\
Minimum angular separation $\Delta R_{\mathrm{min}}$ & 0.05 & 0.05 \\
Pseudorapidity range $|\eta|$ & $0 \leq |\eta| \leq 2.4$ & $0 \leq |\eta| \leq 2.4$ \\
Minimum decay radius $r_{\mathrm{min}}$ & 120 mm (50 mm for piecewise) & 120 mm (50 mm for piecewise) \\
Maximum transverse decay radius $r_{\mathrm{max}}$ & 5 m & 300 mm \\
Maximum longitudinal decay radius $z_{\mathrm{max}}$ & 7 m & 500 mm \\
\bottomrule
\end{tabular}
\begin{tablenotes}
\item[\dag] The piecewise invariant mass cut is a function in the $m_{\ell\ell} \times r_{dv}$ plane, dependent on the position of the HNL decay radius $r_{\mathrm{dv}}$, and invariant mass $m_{\ell\ell}$.
\end{tablenotes}
\end{threeparttable}
\label{tab:cut_table}
\end{table*}

\section{Analysis}
\label{sec:analysis}

The analysis pipeline starts with the Monte Carlo (MC) generation of the process shown in Figure~\ref{fig:feynman} of proton-proton collisions,  $p \, p \rightarrow N \tau^\pm, N \rightarrow \ell^+ \ell^- \vvbar{\tau}$.
%We sample HNL masses $m_N \in \{0.5, 0.75, 1.0,...,20\}$ GeV.
The data generation is performed with MadGraph5\_aMC@NLO \cite{Alwall_2014} (MG) at a center of mass energy of $\sqrt{s}=\SI{13}{TeV}$.
The MC samples provide the four-momenta for all particles involved in the interaction, covering the specified HNL mass range of $0.5 \leq m_N \leq 20$~GeV with the step of $\SI{0.25}{GeV}$.
HNL lifetimes and branching ratios are calculated based on the results of \cite{Bondarenko:2018ptm} and validated against the \texttt{SM\_HeavyN\_NLO} model \cite{Alva:2015, Degrande:2016}.

The macroscopic decay of the HNL is handled in post-processing,\footnote{\href{https://github.com/edtireli/W2HNL}{https://github.com/edtireli/W2HNL}} where lifetimes are drawn from a statistical distribution centered around the theoretical values predicted for the given model parameters ($m_N$, $\Theta_\tau$)~\cite{Bondarenko:2018ptm} and taking into account actual $\gamma$-factor for each of the HNLs in the sample.
Using these sampled lifetimes, the 3D position of the decays is computed in the coordinate system, centered on the interaction point.
%, see Figure~\ref{fig:fiducial_volume}.

The displacement of the decay vertex from the interaction point is determined separately for the transverse and longitudinal directions.

To each di-lepton event we  apply the selection criteria listed in Table~\ref{tab:cut_table} and further discussed in Section~\ref{subsec:selection}.
The event selection criteria, listed in Table~\ref{tab:cut_table}, largely follow those of the displaced vertex analysis in~\cite{ATLAS:2022atq}. The kinematic cuts are standard for this type of analysis, with particular attention given to the invariant mass and decay volume selection criteria, which are discussed in detail in the sections below.

The fiducial decay volumes differ for electrons and muons (see Table~\ref{tab:cut_table}.
In the muon channel, an extended decay volume is used to leverage the muon spectrometer surrounding the inner detector, whereas in the electron channel, the search is restricted to the inner detector region.

% \begin{figure*}[t!]
%     \centering
%     \includegraphics[width=\linewidth]{Validation/side_by_side_plots.png}
%     \caption{Left plot: Full view. Middle plot: $\pm 5 \mathrm{m}$. Right plot: $10.5$ m longitudinal, $500$ mm transverse. All for mass 6 GeV and mixing $\Theta = 10^{-6}$.}
%     \label{fig:cylunder_plot}
% \end{figure*}

The DV reconstruction efficiency is assumed to be 100\%, meaning that any particle meeting the selection criteria is equally likely to be detected in all spatial directions and distances within the modeled decay volume. While it is known that vertex reconstruction efficiency decreases with distance~\cite{ATLAS:2023nze}, properly incorporating this effect outside the ATLAS environment is challenging. Therefore, we have chosen not to include it in our main analysis. In Appendix~\ref{app:atlas_track_reco} we estimate the size of this effect using the efficiency parameterisations of Ref.~\cite{ATLAS:2023nze} and show that it weakens the sensitivity in $\Ut2$ by less than half an order of magnitude, without changing our qualitative conclusions.

% Beyond the DV selection criteria, further criteria are imposed and in general are different for muon and electron channels following [citation for where we get our cuts], and are listed in table \ref{tab:cut_table}.

\subsection{Selection criteria}
\label{subsec:selection}

\subsubsection{Decay volume}
\label{sec:decay_volume}

The detector environment is modeled as a simplified cylindrical model of the ATLAS environment considering $100\%$ detector acceptance and track reconstruction. The minimal decay volume is defined by a spherical criterion, whereby any decay vertex with a radial position below the threshold $r_\text{min}$ is excluded. The outer boundary for the included decay vertices is represented as a cylindrical volume.

\begin{figure*}[p]
    \centering
    \includegraphics[width=0.5\linewidth]{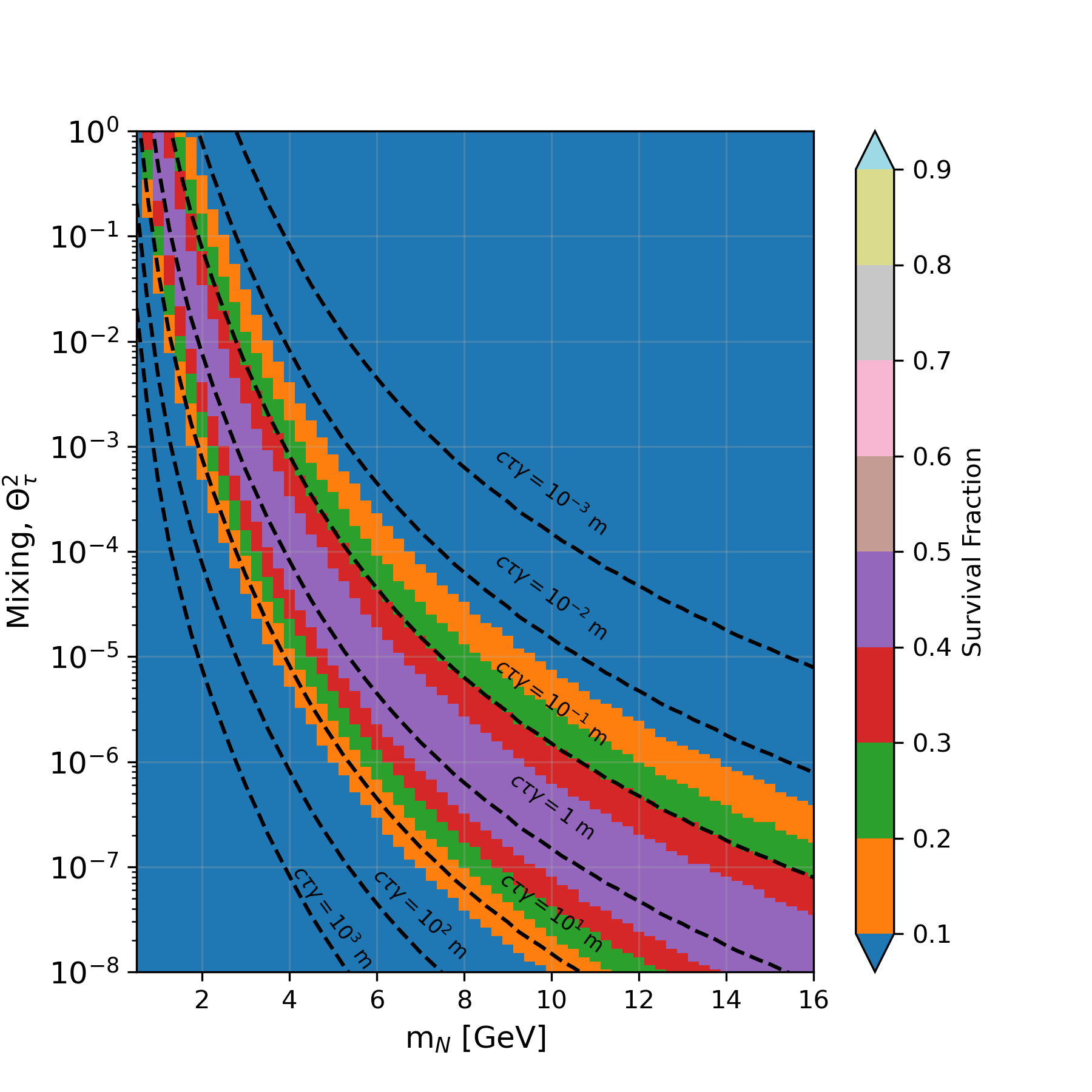}~\includegraphics[width=0.5\linewidth]{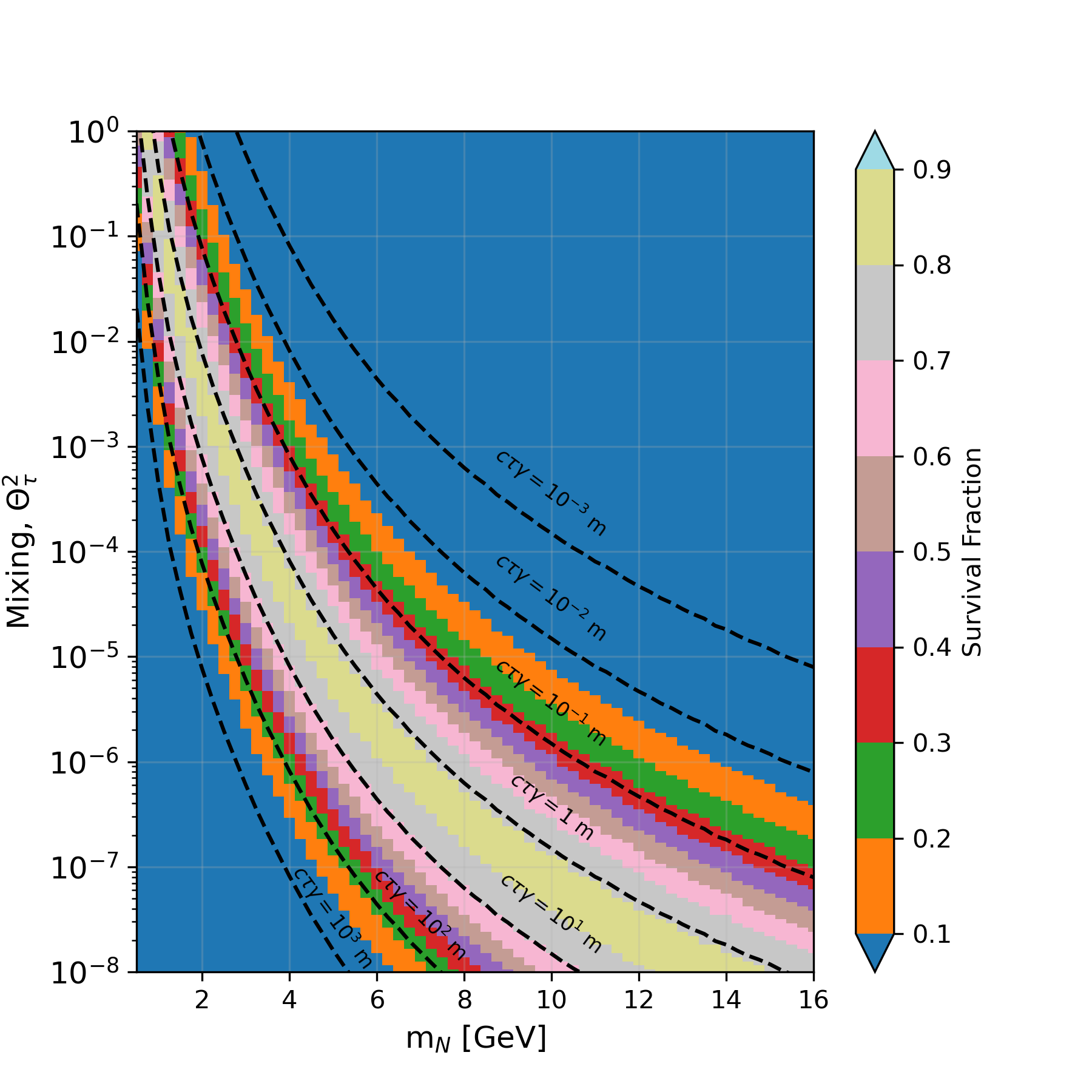}
    \caption{\justifying%
    The fraction of HNLs that decay within the fiducial volume defined by a minimal decay radius, $r_{min}$, and a cylinder with longitudinal, $z_{max}$, and transverse, $r_{max}$, constraints, as specified in Table~\ref{tab:cut_table}. The left panel corresponds to electrons, while the right panel corresponds to muons. The black dashed lines represent lines of constant decay length, $c\tau_N\gamma$ where $\gamma$ is the average Lorentz-factor for the HNL of a given mass.}
    \label{fig:dv_mu}
\end{figure*}

These geometric criteria are detailed in Table~\ref{tab:cut_table}, and the survival efficiencies for the displaced vertex cuts listed in Table~\ref{tab:cut_table} are illustrated in Fig.~\ref{fig:dv_mu}, corresponding to the muon and electron channels, respectively.

\begin{figure*}[p]
    \centering    \includegraphics[width=0.5\linewidth]{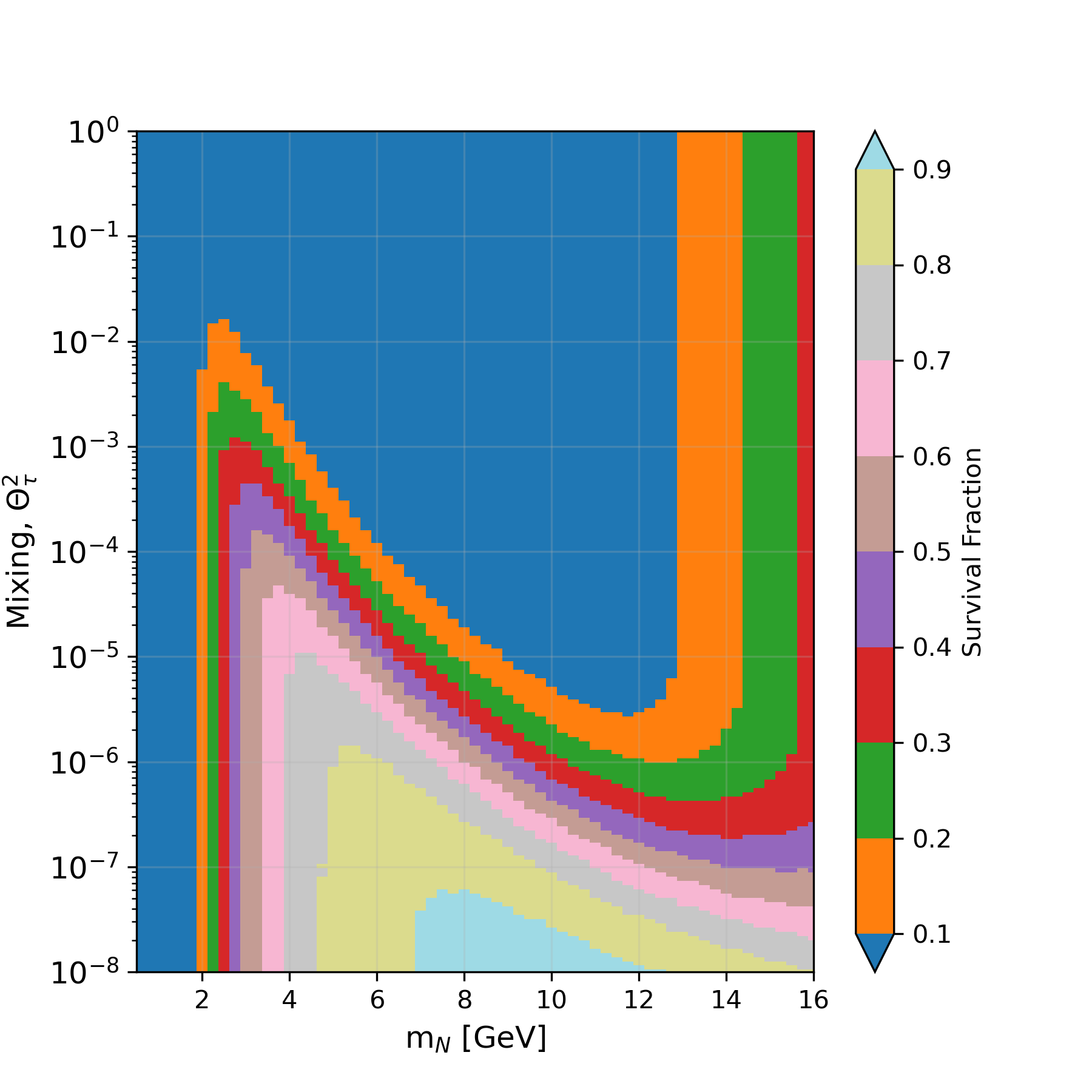}~\includegraphics[width=0.5\linewidth]{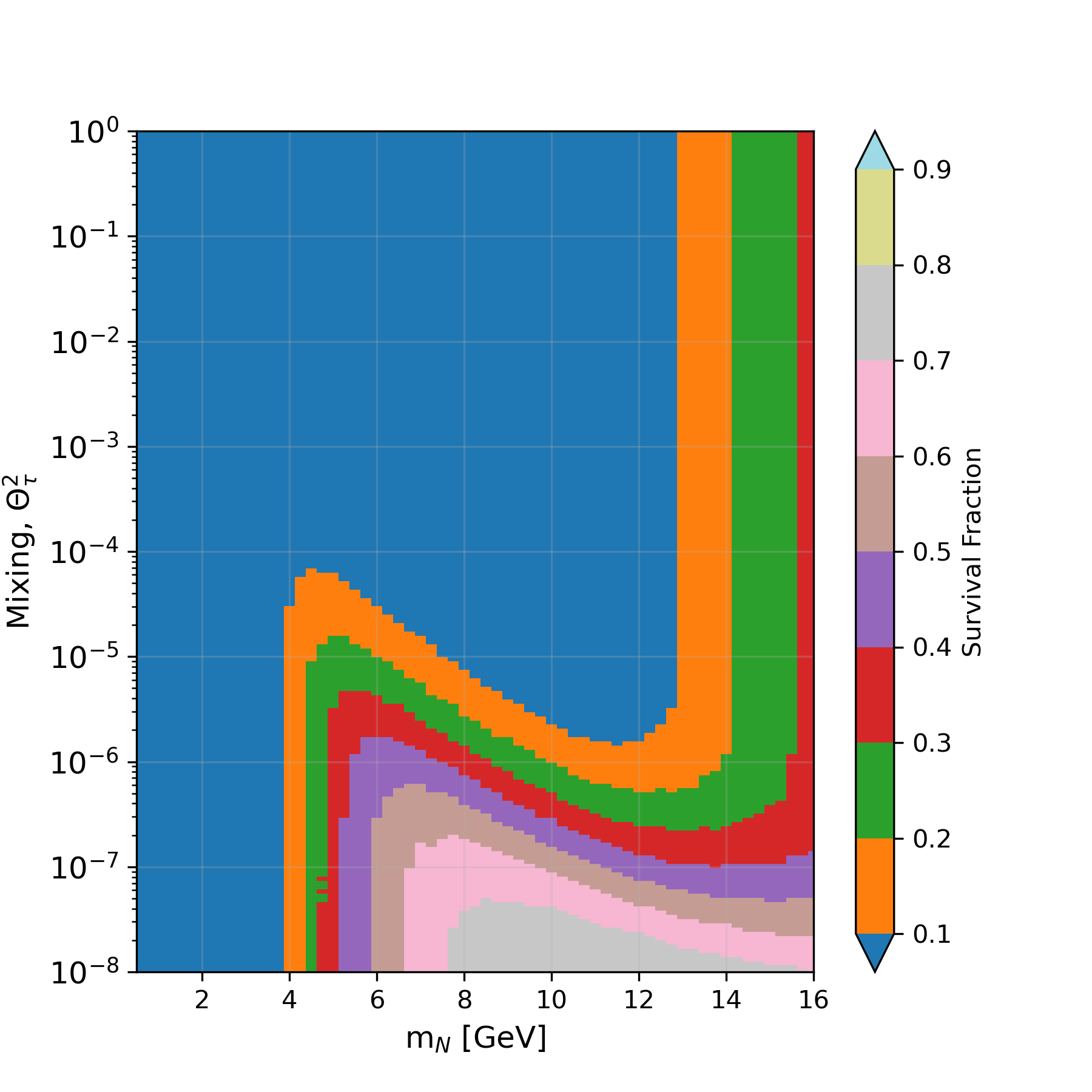}
    \caption{\justifying%
    The survival efficiency of HNLs under the \textit{piecewise} invariant mass selection criteria for the \textit{electron} (left) and \textit{muon} (right) channels. Note that this represents only the efficiency related to the invariant mass of the di-leptons as a function of distance, not the full DV survival probability. To obtain the full radial efficiency, the position of the decay vertices must also be considered, as shown in Figure~\ref{fig:dv_mu}. For further details, see Figure~\ref{fig:inv_DV_survival}.}
    \label{fig:inv_survival}
\end{figure*}
\begin{figure*}[t!]
    \centering
    \includegraphics[width=0.5\linewidth]{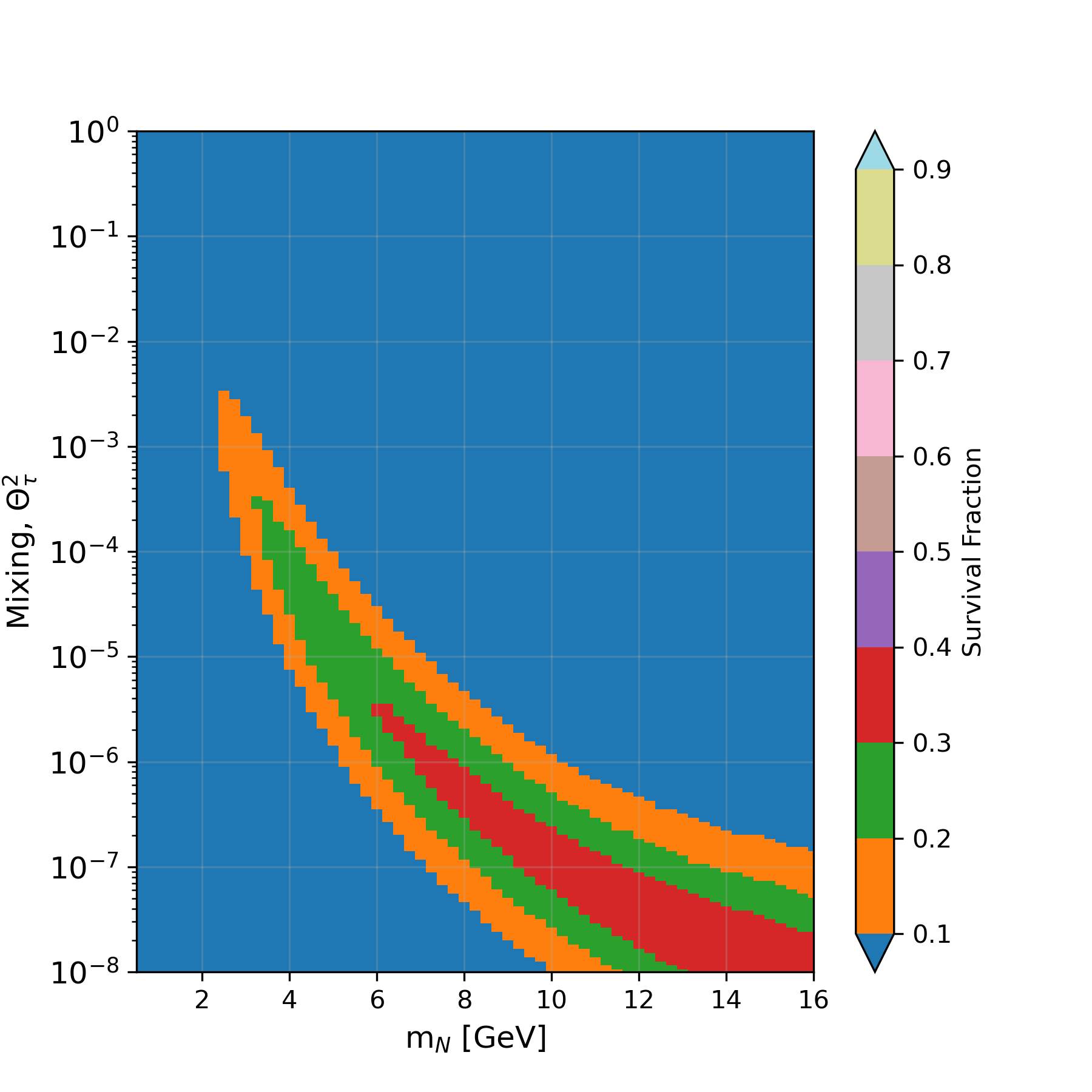}~\includegraphics[width=0.5\linewidth]{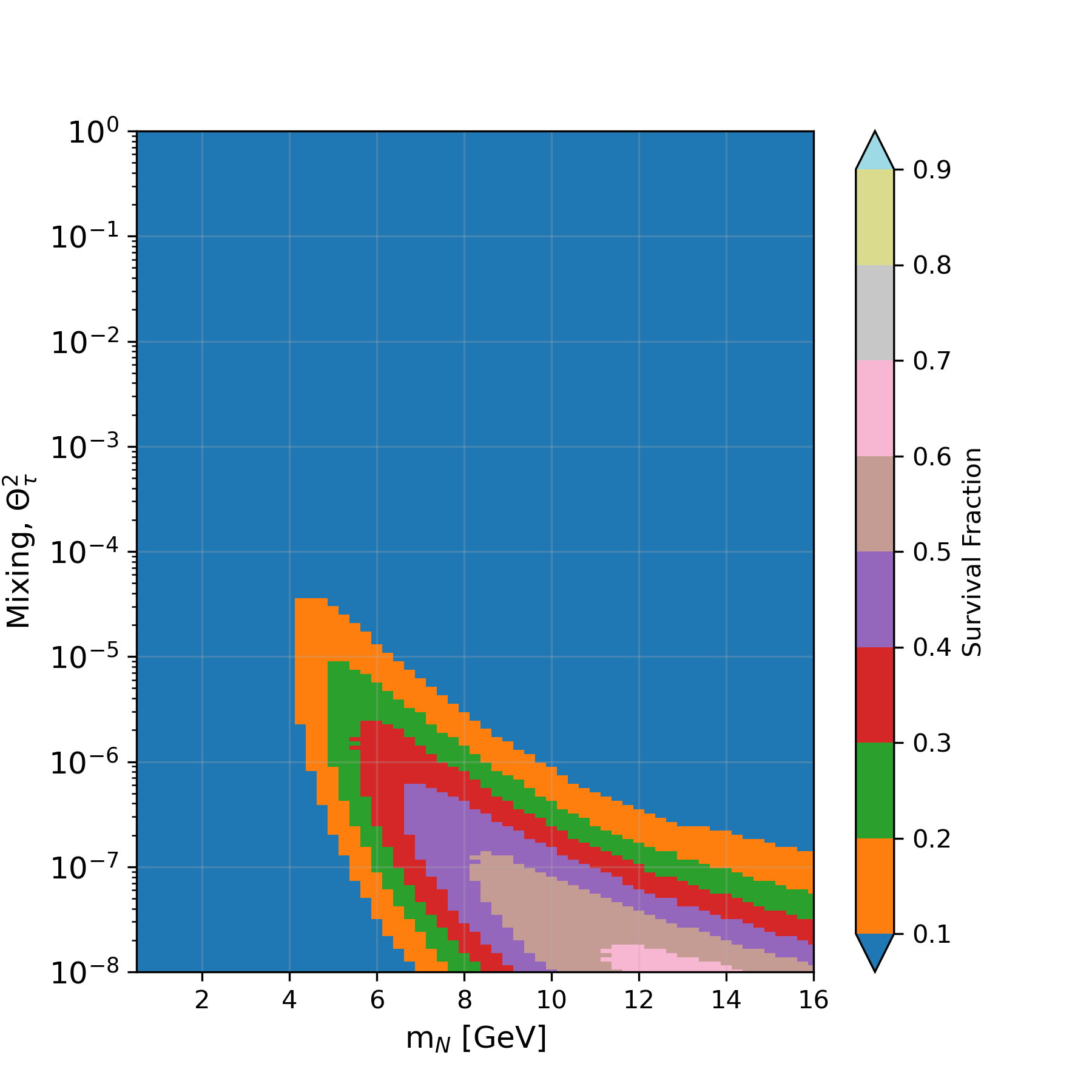}
    \caption{\justifying%
    The survival efficiency of the HNLs from the combined \textit{piecewise} invariant mass selection and DV criteria (composition of Figures~\ref{fig:dv_mu} and~\ref{fig:inv_survival}) for the \textit{electron} (left) and \textit{muon} (right) channel.}
    \label{fig:inv_DV_survival}
\end{figure*}

\subsubsection{Invariant mass}
\label{subsubsec:invmass}

The invariant mass selection criteria have been studied using both the \textit{flat} and \textit{piecewise} approaches for the electron and muon channels. In the \textit{flat} approach, a constant invariant mass cut is applied to all events regardless of the HNL decay position, $r_{DV}$. In contrast, the \textit{piecewise} approach adjusts the mass selection criterion depending on $r_{DV}$.
This method, inspired by the results of \cite{Appelt:2024esk}, allow certain events that would otherwise fail the flat cut to be recovered, see discussion in Section~\ref{sec:background} below.

Figure~\ref{fig:inv_mass_piecewise} illustrates how these two strategies affect events at the lowest $(m_N, \Theta_\tau)$ point in each channel. Green dots indicate events passing the piecewise criteria; red dots indicate those failing. Notably, some events that pass the piecewise cut lie below the flat cut, thereby opening up new regions in parameter space.

Additionally, to demonstrate the overall efficiency impact, we apply cumulative cuts on transverse momentum ($p_T$), pseudorapidity ($\eta$), invariant mass ($m_{\ell\ell}$), and the displaced-vertex criterion ($DV$). Figure~\ref{fig:cumulative} show the progressive effect of these cuts for the di-electron and di-muon channels, respectively.

\begin{figure*}[t!]
    \centering
    \includegraphics[width=0.5\linewidth]{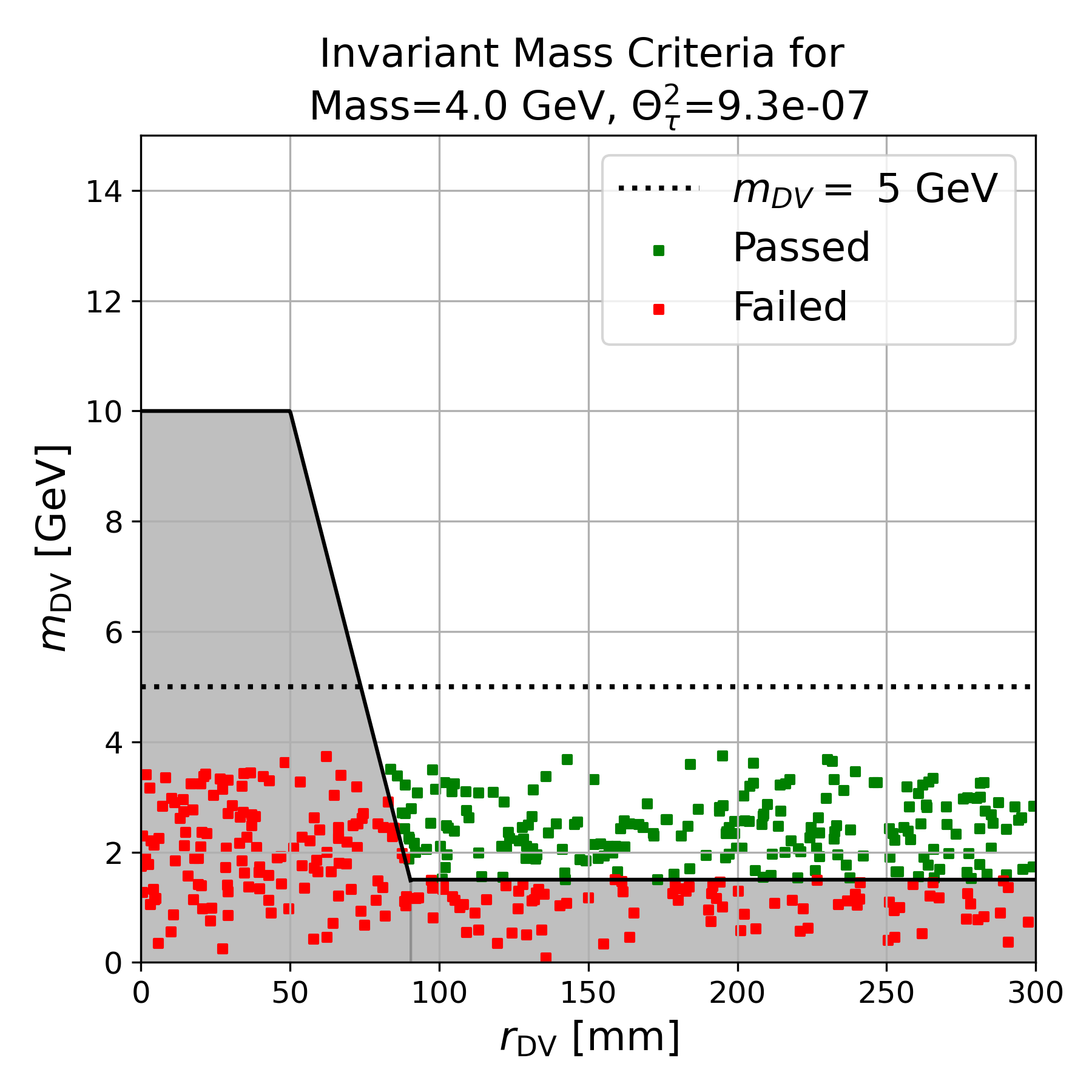}~\includegraphics[width=0.5\linewidth]{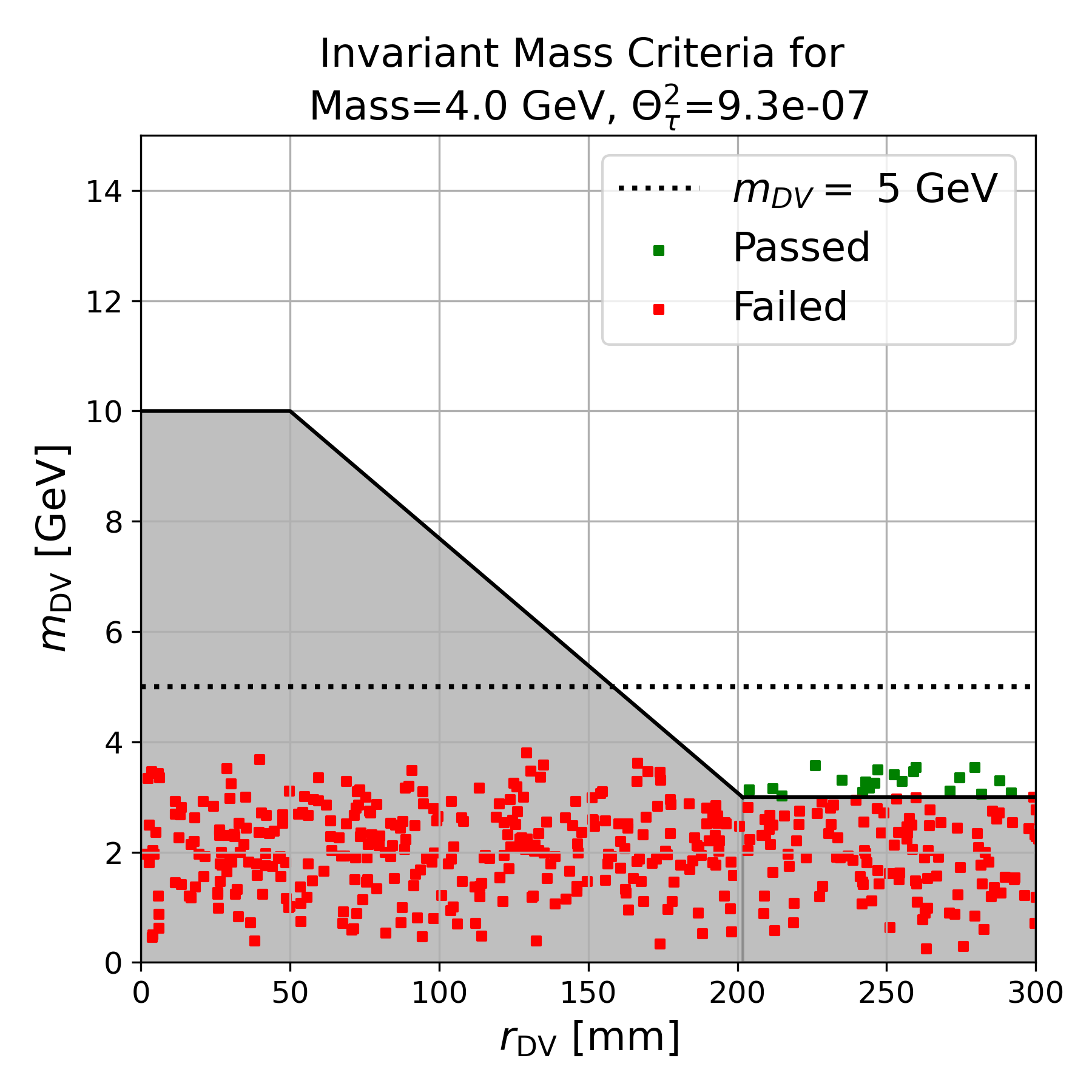}
\caption{\justifying{}\textbf{Efficiency of the flat vs.\ piecewise DV cuts.}
          Invariant mass selection for pairs of \textit{electrons} (left) and \textit{muons} (right) for the HNL with parameters specified in the title of the plot.
          The black line represents the \textit{piecewise} selection approach, while the dotted line corresponds to the \textit{flat} selection.
          Events failing the piecewise cut are shown in red, while those passing are in green.
          Due to the HNL mass constraints, no events survive under the flat invariant mass cut.}
    \label{fig:inv_mass_piecewise}
\end{figure*}

\begin{figure*}[t!]
    \centering
    \includegraphics[width=0.5\linewidth]{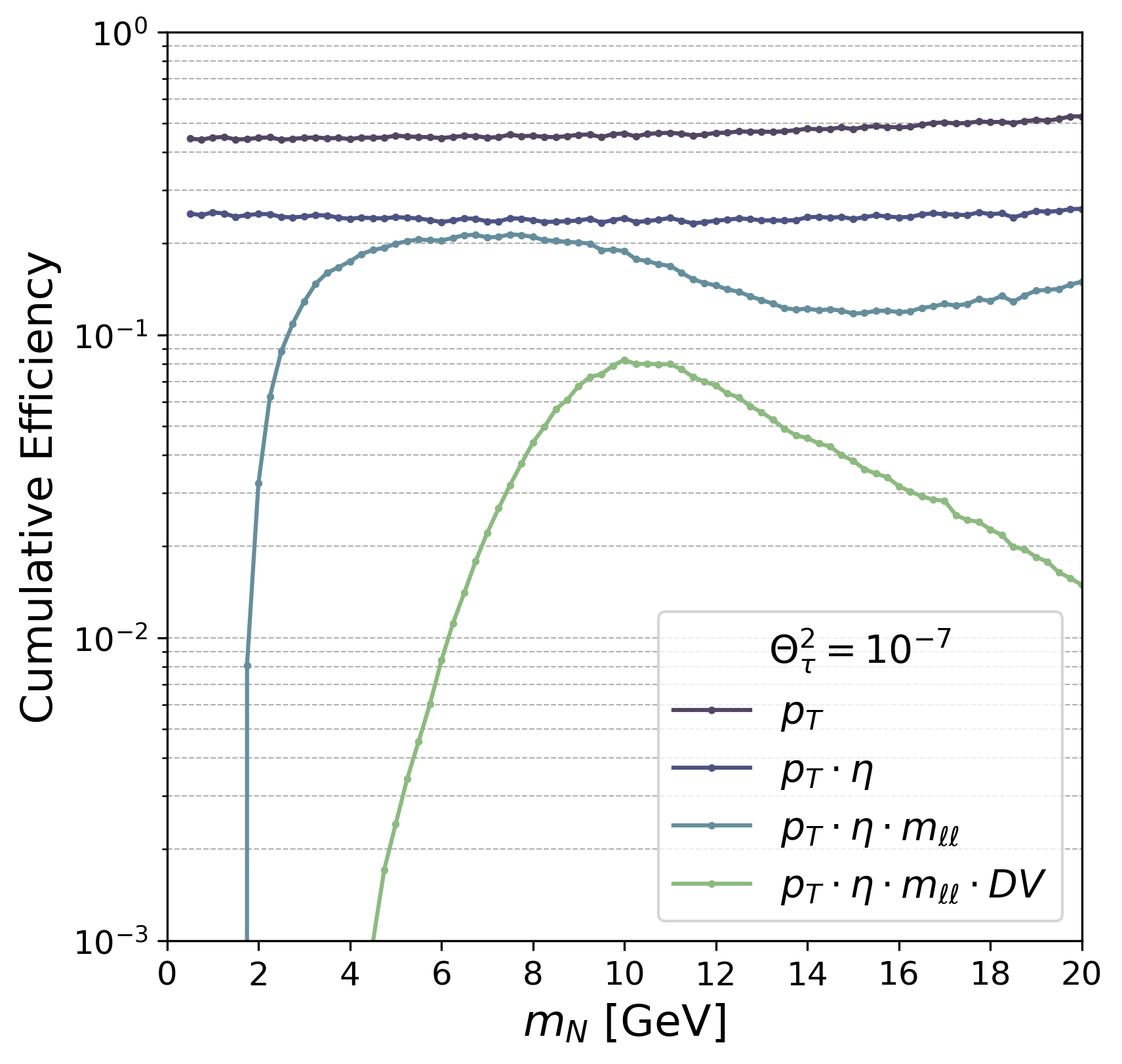}~\includegraphics[width=0.5\linewidth]{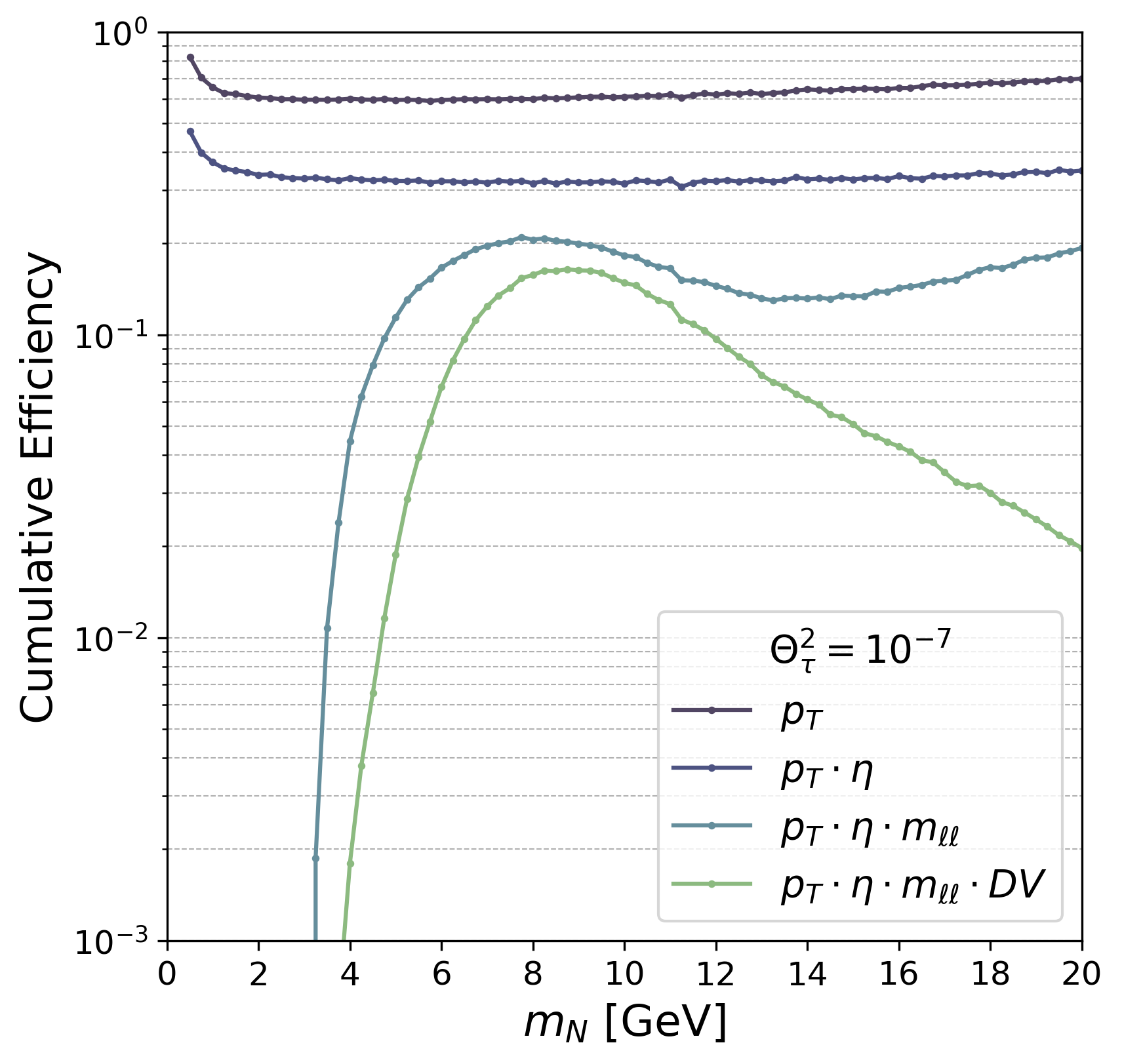}
    \caption{Cumulative effect of the selection criteria ($p_T$, $\eta$, $m_{\ell\ell}$, $DV$)
          applied to daughter leptons: \textit{electrons} (left) and \textit{muons} (right) at $\Theta_\tau^2 = 10^{-7}$.}
    \label{fig:cumulative}
\end{figure*}

\subsection{$\tau$-lepton analysis}
\label{sec:tau_analysis}

% \fxnote{Special care is taken in this analysis to ensure the non-interference of leptons from the prompt $\tau$-lepton. The term is derived from a Monte Carlo analysis of the simulated decay of a $\tau$-lepton with Pythia\cite{pythia}, elaborated in section \ref{sec:tau_analysis}.}

The analysis takes special care to ensure no leptons from the prompt tau may be misidentified as being produced from the HNL,
and as such, any hard leptons must be discarded, leaving only soft leptons of $e, \mu$ flavour and hadronic modes. The production is scaled by this branching fraction derived from a \texttt{Pythia}~\cite{Sjostrand:2014zea} simulation of $10^6$ events of a $13$ TeV center-of-mass proton-proton collision, all $\tau$ decay modes enabled using Pythias sophisticated $\tau$ decay model (\verb|TauDecays:mode = 1|), as well as enabling $\tau$ production from Higgs, $W^\pm$ and $Z^0$ bosons.

\begin{figure}[h!]
    \centering
    \includegraphics[width=1\linewidth]{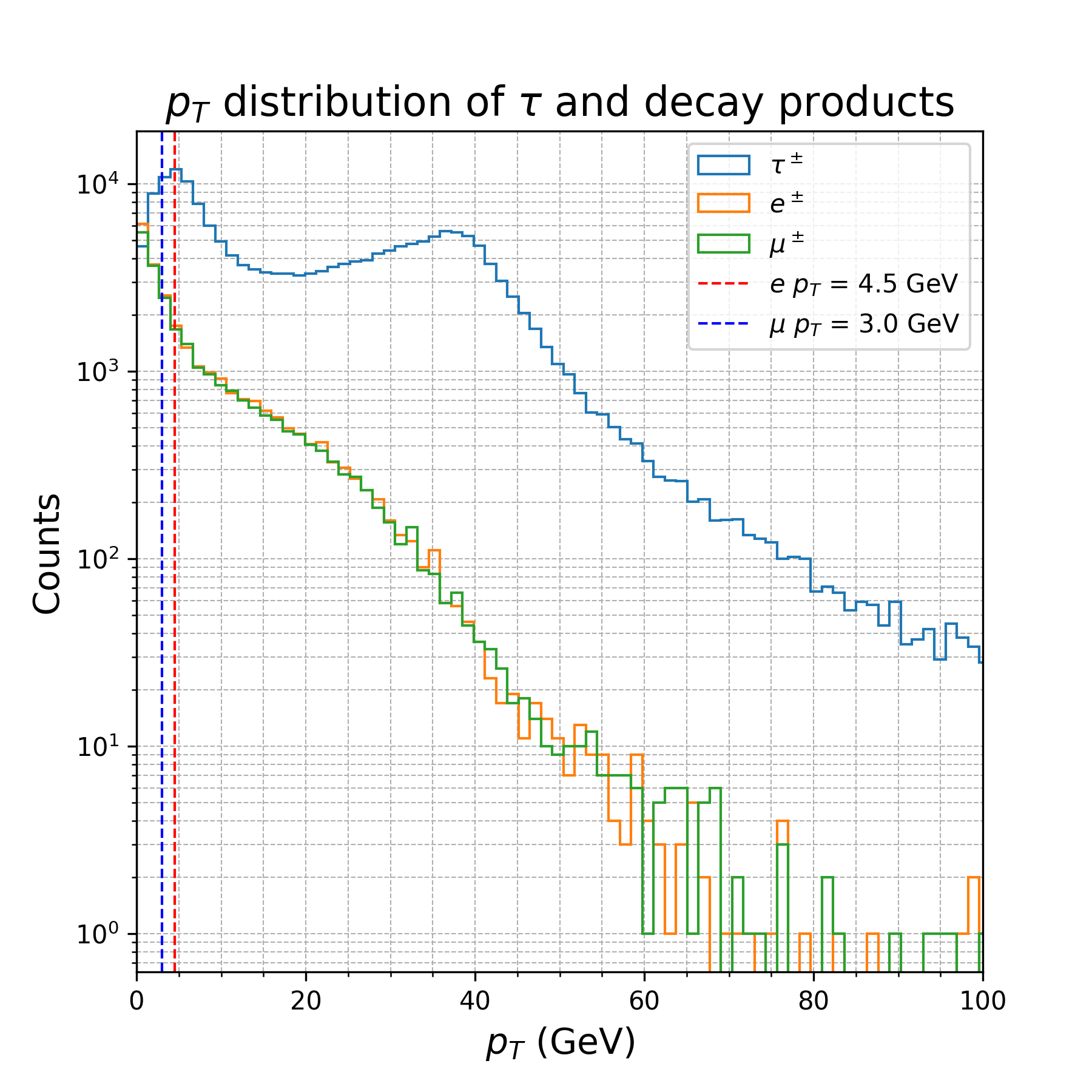}
    \caption{The transverse momenta of the $\tau$-lepton as well as its decay products from the Pythia simulation. The \textit{red} and \textit{blue} dashed line indicates the $p_T$ selection criteria for the electron and muon channels respectively, from table \ref{tab:cut_table}.}
    \label{fig:tau_fractions}
\end{figure}
These fractions are computed as the complement of the theoretical branching to leptonic modes multiplied by the fraction of hard leptons in the simulated analysis, i.e.

\begin{align}
    \Br(\tau \rightarrow X) = \left(1-\text{Br}[\tau \rightarrow \ell \ell \nu]\right) \cdot \left(1 - \frac{N_{\ell}^{\text{soft}}}{N_{\ell}}\right)
\end{align}

This fraction is $0.895$ for the muon channel and $0.912$ for the electron channel and is used in the computation of HNL production in Eq.~\ref{eq:estimate_analytic} and the branchings taken from literature~\cite{ParticleDataGroup:2024cfk}.

\section{Background considerations}
\label{sec:background}

%%%
Our sensitivity estimates so far have been performed under the background-free assumption, which is only partially justified. To assess the background contamination in our signal region, we exploit the fact that it coincides with the \textit{validation region} used by the ATLAS collaboration~\cite{PhysRevLett.131.061803}, which contains events with a displaced lepton pair but no prompt lepton. The results are presented in detail in the Ph.D.\ thesis~\cite[Section 5.7]{Appelt:2024esk}.

The dominant sources of background are the decays of meta-stable particles and accidental track crossings involving cosmic muons. The heavy flavour background in the $\mu\mu$ channel was eliminated in~\cite{PhysRevLett.131.061803} by imposing the invariant mass cut $\mdv \ge \SI{5.5}{GeV}$. The accidental track crossing background is largely irreducible, but can be assessed by comparing same-sign and opposite-sign muon pairs, which are approximately equal in the case of track crossings. Additionally, in the $ee$ channel, interactions of particles with the detector material and photon conversions can create fake displaced vertices close to high-density regions.

With this background picture in mind, we implemented a \textit{piecewise} selection in the $(\mdv, \rdv)$ plane, inspired by the ATLAS $ee$ channel veto and extended in this work to the $\mu\mu$ channel. The shape of the selection region is shown in dark grey in Figure~\ref{fig:inv_mass_piecewise}. We compared this selection with a \textit{flat} selection requiring $M_{\mathrm{inv}}^{\mathrm{min}} > \SI{5}{GeV}$ and positions of DV within the fiducial cylinder, following the ATLAS approach~\cite{PhysRevLett.131.061803}. While the flat selection would eliminate the signal entirely, the piecewise selection allows us to preserve part of the signal while efficiently reducing the background.

For the material-induced background in the $ee$ channel, a 3D veto near the high-density regions was imposed in~\cite{PhysRevLett.131.061803}, leading to an approximately $42\%$ reduction of the fiducial volume~\cite{PhysRevLett.131.061803,Appelt:2024esk}. We do not propagate this reduction in our work, as we cannot properly fold it with the density of DVs without having the detailed map of the detector materials. It should be noted that Figure~5.28 in~\cite{Appelt:2024esk} does not exhibit any noticeable increase in $ee$ background around material layers, which further justifies our treatment.

%%%

Further background suppression is achieved through additional selection criteria:
\begin{itemize}
    \item The majority of the explored HNL parameter space satisfies $N \geq 3$ events in \textit{both} electron and muon channels, as shown in Figure~\ref{fig:main_result_channels}. This cross-channel requirement serves as an effective background rejection criterion.
    \item We have not yet utilized the presence of a prompt $\tau$-lepton in the signal, which could further suppress background contributions.
    \item Figure~\ref{fig:main_result_Nevents} demonstrates that the piecewise selection criteria allow for regions with $N \gg 1$ events, increasing the robustness of the analysis.
\end{itemize}

These considerations suggest that searches for HNLs decaying into a displaced lepton pair and a prompt $\tau$-lepton can be made effectively background-free.

\section{Discussion}
The results presented in this work demonstrate that displaced vertex searches in the process $pp \to W \to \tau N$, with $N \to \ell^+ \ell^- \nu_\tau$, can  probe previously unexplored regions of the $\nu_\tau$–HNL mixing parameter. % space using Run~2 data.
The adopted radial displacement requirement, $\rdv \geq 100~\mathrm{mm}$, effectively suppresses backgrounds from heavy flavor decays while preserving signal sensitivity. Moreover, the introduction of a \emph{piecewise} invariant mass selection in the $(\mdv, \rdv)$ plane substantially improves efficiency relative to a flat invariant-mass cut, particularly in regions where the decay topology would otherwise fall below fixed thresholds~\cite{PhysRevLett.131.061803,Appelt:2024esk}.

The explored topology primarily populates decay radii around the first SCT layers of the ATLAS detector, highlighting the strong physics case for extending the capabilities of large radius tracking (LRT) to larger displacements and improving reconstruction of displaced electrons~\cite{ATLAS:2023nze}. Such detector level improvements would directly enhance the discovery potential of searches targeting long-lived states originating from $W$-boson decays, including HNLs coupled to the $\tau$-flavor.

A positive signal in this channel would open a new experimental window into the sterile neutrino sector, linking neutrino mass generation, baryogenesis, and possible dark-sector production mechanisms.
The absence of a signal will contribute to further exploration of the
parameter space for GeV-scale $\nu_\tau$–dominated heavy neutral leptons, providing valuable constraints for models that attempt to connect neutrino physics and other BSM phenomena.
% Conversely, the absence of such a signal will allow exclusion of the most theoretically motivated parameter space for GeV-scale $\nu_\tau$–dominated heavy neutral leptons, providing valuable constraints for models that attempt to connect neutrino physics and dark matter~\cite{Boyarsky:2018tvu}.

\section{Conclusion and outlook}
\label{sec:conclusion}

Heavy neutral leptons remain a compelling extension of the Standard Model, capable of addressing several open questions in fundamental physics, including the origin of neutrino masses, the generation of the baryon asymmetry of the Universe, and even the nature of dark matter. A particularly promising experimental signature of HNLs is the appearance of displaced vertex  events, originating from long-lived particles decaying within the detector. While LHC searches have so far focused on HNLs mixing with electron or muon neutrinos, direct constraints on HNLs coupled to the tau-neutrino flavor remain underexplored.

In this work, we demonstrated that existing DV searches (such as \cite{ATLAS:2022atq} or \cite{CMS:2022fut}) can be naturally extended to probe HNLs with dominant $\nu_\tau$ mixing. We considered the process $pp \to \tau^\pm N$, with the HNL decaying via $N \to \ell^+ \ell^- \nu_\tau$, and studied the final states $\mu^+\mu^-$, $e^+e^-$, and their combination. Using Monte Carlo simulations and a simplified ATLAS detector geometry, we estimated the signal acceptances and projected sensitivities for integrated luminosities of $\SI{139}{\per\femto\barn}$, $\SI{300}{\per\femto\barn}$, and $\SI{3000}{\per\femto\barn}$.

A central feature of such an analysis remains the suppression of Standard Model backgrounds.
The dominant background originates from decays of heavy flavor hadrons, populating the region of small invariant mass ($\mdv \lesssim \SI{5.5}{GeV}$) and short displacement ($\rdv \lesssim \SI{50}{mm}$). We evaluated two selection strategies in the $\mdv$--$\rdv$ plane: a simple flat invariant mass cut, and a more refined piecewise cut that adapts the $\mdv$ threshold based on the DV radius. The latter, inspired by existing ATLAS strategies in the $ee$ channel, was extended in this work to the $\mu\mu$ channel.
We argued that these optimized cut strategies may allow for efficient background suppression while maintaining high signal acceptance.

We found that, even at Run 2 luminosities, parts of parameter space previously untested by experiments such as DELPHI and BEBC become accessible. Future data-taking campaigns, particularly at the HL-LHC, could improve these bounds by one order of magnitude or more. The combined channel notably enhances robustness and offers additional leverage for background rejection. Overall, our study shows that with optimized selection criteria and realistic background modeling, DV-based searches can provide meaningful constraints on HNLs mixing with the tau-neutrino flavor, unachievable by other experiments.

Further progress will require full experimental implementation, including detector-specific effects, reconstruction efficiencies, and potential residual backgrounds beyond those visible in validation regions. Nevertheless, we believe that our work provides a clear framework and concrete benchmarks for extending DV searches into this less-explored sector of HNL parameter space. A reinterpretation of existing LHC data along these lines would be both feasible and timely.

\section{Acknowledgments}
The authors would like to thank Mads Mølbak Hyttel, who contributed to the early stage of this analysis.
We also thank C.~Appelt and A.~Soffer for reading the manuscript and providing valuable comments concerning the efficiency of DV detection.

\section{Data availability}
The simulation and post-processing code used to generate the Monte Carlo
samples, apply the selection criteria and derive the sensitivity contours
reported in this work, together with the resulting datasets, are openly
available in Ref.~\cite{W2HNL}.

\appendix

\section{Impact of the DV reconstruction efficiency}
\label{app:atlas_track_reco}

The headline sensitivity contours in Figures~\ref{fig:main_result_Nevents}
and \ref{fig:main_result_channels}, as well as the survival efficiencies of
Section~\ref{sec:analysis}, were obtained under the assumption that the
displaced-vertex reconstruction efficiency is $100\%$ everywhere within the
fiducial volume. As discussed at the end of Section~\ref{sec:analysis} and
in line with the original ATLAS analysis~\cite{ATLAS:2022atq}, this is
known to be an idealisation: vertex- and track-reconstruction efficiency
decreases with the distance of the displaced vertex from the interaction
point, primarily because tracks with large transverse impact parameter
$|d_0|$ and tracks originating beyond the first pixel layers are harder to
reconstruct.

In this Appendix we estimate the size of this effect. We use the dedicated ATLAS large-radius tracking
performance study of Ref.~\cite{ATLAS:2023nze}, whose Fig.~7 reports the
per-track reconstruction efficiency for the $W \to N\ell$ topology as a
function of the transverse impact parameter $|d_0|$ and the transverse
production radius $R_\text{prod}$. We digitise both curves and treat the
two efficiencies as factorised and independent per track,
\begin{equation}
\label{eq:atlas_track_reco}
  \epsilon_\text{track}
  \;=\;
  \epsilon_{d_0}(|d_0|) \,\cdot\, \epsilon_{R}(R_\text{prod}) .
\end{equation}
For every displaced lepton in our Monte Carlo sample we then reject the
track stochastically: an independent uniform random variable $u \in [0,1)$
is drawn and the track is kept if $u < \epsilon_\text{track}$. An event
survives only if \emph{both} displaced tracks pass the per-track draw. In
the limit of large statistics this procedure recovers the factorised
acceptance probability of Eq.~(\ref{eq:atlas_track_reco}) while preserving
the binary, event-by-event signature used throughout the rest of the
analysis.

We stress that Eq.~(\ref{eq:atlas_track_reco}) is an approximation: the
ATLAS curves were extracted for one specific BSM topology and detector
configuration, the factorisation between $|d_0|$ and $R_\text{prod}$ is not
exact, and our analysis is not performed within the ATLAS software
environment. We therefore view the results below as a conservative
\emph{indication} of the magnitude of the effect rather than a
quantitative replacement of the main results.

\begin{figure*}[!ht]
    \centering
    \subfloat[$\mathcal{L} = \SI{139}{\femto\barn^{-1}}$, piecewise]{\includegraphics[width=0.32\textwidth]{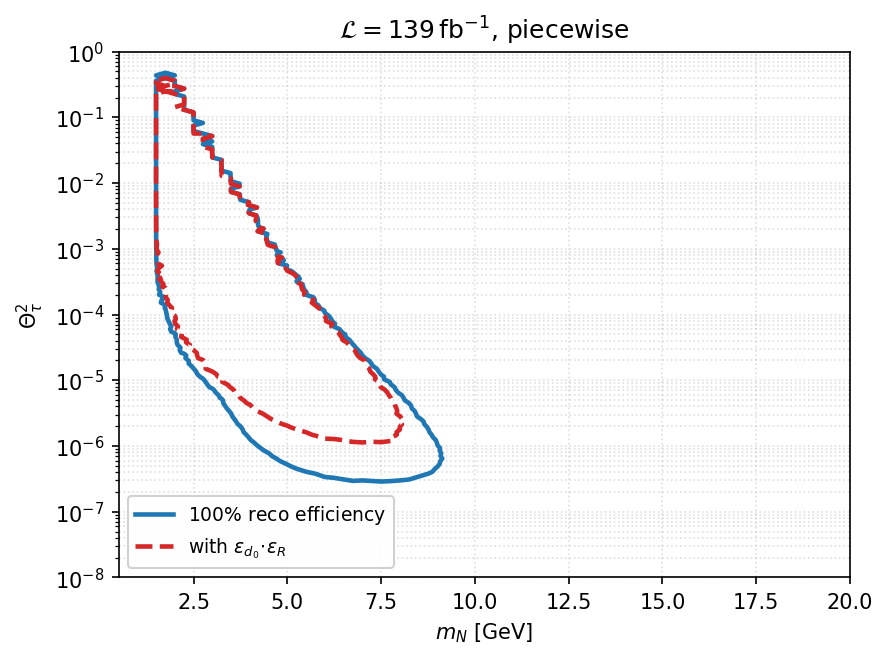}}~
    \subfloat[$\mathcal{L} = \SI{300}{\femto\barn^{-1}}$, piecewise]{\includegraphics[width=0.32\textwidth]{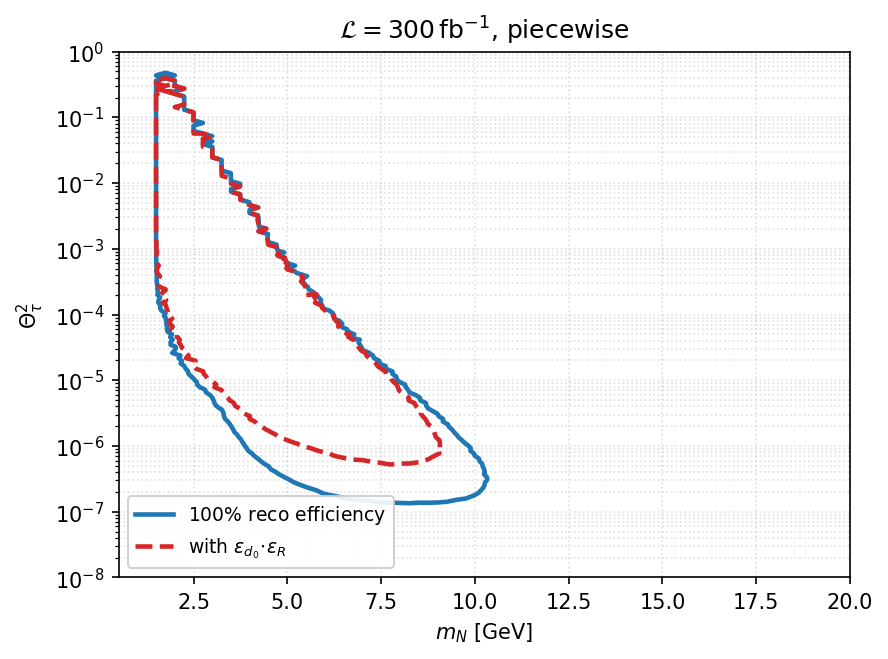}}~
    \subfloat[$\mathcal{L} = \SI{3000}{\femto\barn^{-1}}$, piecewise]{\includegraphics[width=0.32\textwidth]{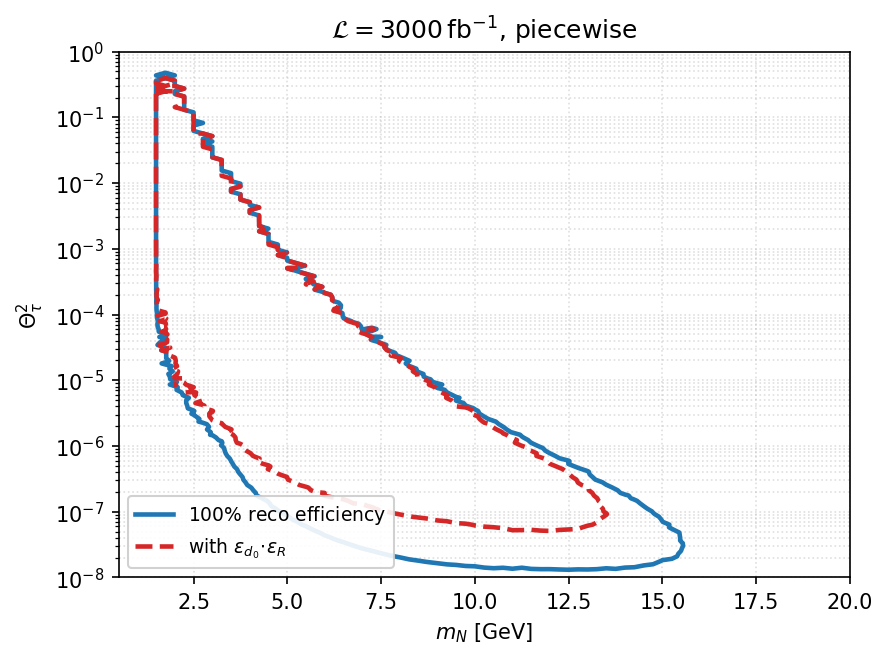}}
    \caption{\justifying%
    Effect of the per-track ATLAS reconstruction efficiency on the
    sensitivity contours, for the three integrated luminosities considered
    in the main text and the piecewise invariant-mass selection. Solid
    contours correspond to the $100\%$-efficiency assumption used in the
    main analysis; dashed contours show the same selection with the
    per-track efficiency factor of Eq.~(\ref{eq:atlas_track_reco}) applied.
    The combined-channel ($ee+\mu\mu$) acceptance is reduced by a factor
    $\sim 0.61$ at peak, corresponding to a shift of the $N\ge 3$ contour
    in $\Ut2$ by a factor $\sim 1.3$, with the largest impact at small
    mixing angles where the typical decay length pushes the displaced
    vertices into the low-efficiency region of Ref.~\cite{ATLAS:2023nze}.}
    \label{fig:atlas_sensitivity_shift}
\end{figure*}

The resulting impact on the signal acceptance is summarised in Table~\ref{tab:atlas_eff_impact}, and its effect on the $N\ge 3$ sensitivity contours is shown in Figure~\ref{fig:atlas_sensitivity_shift}, where we overlay the contours of Figs.~\ref{fig:main_result_Nevents}--\ref{fig:main_result_channels} with and without the per-track efficiency factor of Eq.~(\ref{eq:atlas_track_reco}). The suppression is stronger in the $\mu\mu$ channel because its extended fiducial volume ($r_\text{max} = \SI{5}{m}$, $z_\text{max} = \SI{7}{m}$) probes larger displacements, where the $\epsilon_R$ digitisation of Ref.~\cite{ATLAS:2023nze} drops below $\sim 0.6$. Because the expected event count scales as $\Theta_\tau^4$ at short lifetimes, an acceptance ratio $r$ translates into a shift of the $N\ge 3$ contour by a factor $1/\sqrt{r}$ in $\Ut2$. In the combined channel this shift is a factor of $\sim 1.3$--$1.5$, i.e.\ less than half an order of magnitude, and does not change the qualitative conclusions of Section~\ref{sec:result}: the parameter space currently constrained only by DELPHI~\cite{DELPHI:1996qcc} and BEBC~\cite{Barouki:2022bkt} remains accessible already at Run~2 luminosities, while the projected high-luminosity reach is reduced but still extends well beyond the present limits.

\begin{table}[!ht]
\centering
\caption{\justifying%
Impact of the per-track reconstruction-efficiency factor of
Eq.~(\ref{eq:atlas_track_reco}) on the mean signal acceptance. The second
column gives the ratio of acceptances with and without the efficiency
applied; the third column gives the corresponding shift of the $N\ge 3$
contour in $\Ut2$, computed as $1/\sqrt{r}$ (see text).}
\label{tab:atlas_eff_impact}
\begin{tabular}{lcc}
\hline\hline
Channel / selection & ratio $r$ & shift in $\Ut2$ \\
\hline
$ee$, piecewise                    & 0.38 & $\approx 1.6\times$ \\
$ee$, flat                         & 0.30 & $\approx 1.8\times$ \\
$\mu\mu$, piecewise                & 0.20 & $\approx 2.3\times$ \\
$\mu\mu$, flat                     & 0.20 & $\approx 2.2\times$ \\
\hline
combined ($ee+\mu\mu$), piecewise  & 0.61 & $\approx 1.3\times$ \\
combined ($ee+\mu\mu$), flat       & 0.43 & $\approx 1.5\times$ \\
\hline\hline
\end{tabular}
\end{table}

A more complete treatment would require running our signal samples through
the full ATLAS simulation chain, including the dedicated large-radius
tracking algorithms described in Refs.~\cite{ATLAS:2023nze, ATLAS:2022atq},
which is beyond the scope of the present generator-level study. The
estimate of this Appendix shows, however, that the omission of the
radial-dependent reconstruction efficiency in the main analysis introduces
a correction at the factor-of-a-few level, and motivates a future
experimental implementation of the search proposed here.

\bibliography{bibliography}

\end{document}